\documentclass[preprint]{aastex}
\begin{document}
%%%\begin{center}
\title{~~\\ ~~\\ Multi-Frequency Optical-Depth Maps and the Case for Free-Free Absorption in Two Compact Symmetric Radio Sources: the CSO candidate $J1324+4048$ and the CSO $J0029+3457$}

\author{J.\ M.\ Marr\altaffilmark{1}, T.\ M.\ Perry\altaffilmark{2}, 
J.\ Read\altaffilmark{1}, G.\ B.\ Taylor\altaffilmark{3}, and A.\ O.\ Morris\altaffilmark{1}}

\altaffiltext{1}{Union College, Schenectady, NY  12308, USA 
$<$marrj@union.edu$>$}

\altaffiltext{2}{University of Wisconsin, Madison, WI 53706, USA}

\altaffiltext{3}{University of New Mexico, Albuquerque, NM 87131, USA; 
Greg Taylor is also an
Adjunct Astronomer at the National Radio Astronomy Observatory.}

%\received{year Month date}
%\accepted{year Month date}

\slugcomment{Accepted by the Astrophysical Journal: 21-November-2013}

\begin{abstract}

We obtained dual-polarization VLBI observations at six frequencies of the compact symmetric object $J0029+3457$ and the CSO candidate $J1324+4048$.  By comparing the three lower-frequency maps with extrapolations of the high frequency maps we produced maps of the optical depth as a function of frequency. The morphology of the optical-depth maps of $J1324+4048$ is strikingly smooth, suggestive of a foreground screen of absorbing gas.  The spectra at the intensity peaks fit a simple free-free absorption model, with $\chi_\nu^2 \approx 2$, better than a simple synchrotron self-absorption model, in which $\chi_\nu^2 \approx 3.5 - 5.5$.  We conclude that the case for free-free absorption in $J1324+4048$ is strong.  The optical-depth maps of $J0029+3457$ exhibit structure, but the morphology does not correlate with that in the intensity maps.  The fit of the spectra at the peaks to a simple free-free absorption model yields $\chi_\nu^2 \approx 1$, but since the turnover is gradual the fit is relatively insensitive to the input parameters.  We find that free-free absorption by a thin amount of gas in $J0029+3457$ is likely, but not definitive.  One compact feature in $J0029+3457$ has an inverted spectrum even at the highest frequencies.  We infer this to be the location of the core and estimate an upper limit to the magnetic field of order 3 Gauss at a radius of order 1 pc.  In comparison with maps from observations at earlier epochs, no apparent growth in either $J1324+4048$ or $J0029+3457$ is apparent, with upper limits of 0.03 and 0.02 mas yr$^{-1}$, corresponding to maximum linear separation speeds of 0.6c and 0.4c.

\end{abstract}


\section{INTRODUCTION}

The radio sources $J1324+4048$ 
(RA=13$^h$ 24$^m$ 12.09$^s$, Dec=40$^\circ$ 48' 11''.8 (J2000); z=0.496 (Vermeulen et al.\ 1996)) 
and $J0029+3457$ 
(RA=00$^h$ 29$^m$ 14.24$^s$, Dec=34$^\circ$ 56' 32''.3 (J2000); z=0.517 (Zensus et al.\ 2002)) 
have spectra characteristic of a class of extragalactic radio source referred to as Gigahertz-Peaked Spectrum (or GPS) sources, which are defined by large radio luminosities, compact structure, and spectra that turn over at GHz frequencies with steep spectra on either side of the peak (O'Dea et al.\ 1991).  An extensive discussion of GPS sources is provided in a review by O'Dea (1998).  

These two sources can also be classified by their morphology.  There are two general morphological types of GPS sources -- core-jet structures, generally associated with quasars and at high redshifts, and compact double-lobed structures, generally associated with galaxies at lower redshifts (O'Dea et al.\ 1991; Conway et al.\ 1994; Stanghellini et al.\ 1997).  The latter group has garnered much interest because of their appearance as miniature versions of the classical double-lobed radio galaxies.  Studies of these sources generally refer to them as ``Compact Symmetric Objects'' (or CSOs), defined by Wilkinson et al.\ (1994) as radio sources with symmetric compact structure containing two steep-spectra, lobe-like features bracketing a compact, flat or inverted spectrum, core-like component.  In many cases, though, the core component is undetected and so some such sources may be called ``compact doubles'' (CDs) (e.g.\ Phillips \& Mutel 1982; Mutel et al.\ 1985) or ``CSO candidates'' (e.g.\ An \& Baan 2012).  The two sources we discuss here contain two steep-spectra, lobe-like components (Henstock et al.\ 1995).  The source $J0029+3457$ has previously been identified as a CSO (Augusto et al.\ 2006) while $J1324+4048$ has been identified as a CSO candidate (An et al.\ 2012).  

\subsection{Young Sources or Frustrated Jets?}

Many authors have suggested that CSOs are young precursors to the large radio galaxies (e.g.\ Phillips \& Mutel 1982; Carvalho 1985; de Young 1993; Fanti et al.\ 1995; Murgia 2003), while others have proposed that the growth of the jets is frustrated by interaction with dense ambient gas in the nuclei of the host galaxies (van Breugel 1984; Gopal-Krishna \& Wiita 1991).  Other models involve both youth and frustration as the jets grow slowly in the early stages as the jets work their way through denser gas at smaller radii (e.g.\ Begelman 1999; Bicknell et al.\ 1997; Snellen et al.\ 2000; Tinti \& De Zotti 2006; Kawakatu et al.\ 2008). 

The GPS sources with CSO morphologies are generally considered as an extension of the Compact Steep Spectrum (or CSS) sources, whose spectra turnover below 1 GHz and which have intermediate sizes, larger than CSOs but significantly smaller than the larger radio galaxies (or LRGs).  In a statistical study of CSS sources, Fanti et al.\ (1990) found a strong anti-correlation between the linear size, LS, and turnover frequency, $\nu_{\rm turnover}$.  In the youth model, GPSs grow into CSSs and then into the LRGs.  

In strong support of the youth model, multi-epoch VLBI studies of many CSOs have revealed outward motions of features, with growth rates of $\sim$0.1 c, implying ages of the radio-emitting structure from many hundreds to a few thousand years (Owsianik \& Conway 1998; Taylor et al.\ 2000; Pihlstr{\"o}m et al.\ 2003; Gugliucci et al.\ 2005; Nagai et al.\ 2006).  Other age estimates obtained by applying energy considerations (Readhead et al.\ 1996) and by fitting the high-frequency spectral breaks to continuous injection models (Murgia et al.\ 1999) have yielded values astonishingly close to the measured kinematic ages.  These agreements also imply that equipartition of energy exists in these sources.  

Statistical studies revealing an overabundance of CSOs and CSSs relative to LRGs (O'Dea \& Baum 1997) presented challenges to the simple youth model, leading to the suggestion that a fraction of the smaller sources turn off after a relatively short period (Readhead et al.\ 1994).  Some models, though, successfully fit the population statistics by considering differing luminosity evolutions due to intermittency of nuclear activity (Reynolds \& Begelman 1997; Begelman 1999) or to effects of a varying ambient gas density (O'Dea \& Baum 1997; Snellen et al.\ 2000; Perucho \& Mart{\'i} 2002).  More recently, An and Baan (2012) measured the number of CSOs as a function of the linear size and of the kinematic age and found the numbers to decrease in both cases, suggesting that a significant fraction of CSO's do indeed die out on a time scale of $\sim$ 1000 years.

\subsection{FFA vs.\ SSA}

The radio-frequency emission in GPS sources above the spectral turnover is generally accepted as optically-thin synchrotron emission with the turnover of the spectra resulting from absorption at lower frequencies.  However, the absorption mechanism responsible is still an unsettled issue.  On one hand, ionized gas in the host galaxies, with densities and columns depths typical of the narrow-emission-line regions (or `NELR'), would cause free-free absorption (or `FFA') with optical depths of the right magnitude to cause spectral turnovers at GHz frequencies (van Breugel 1984; Bicknell et al.\ 1997; Begelman 1999; Stawarz et al.\ 2008; Osterero et al.\ 2010), while, on the other hand, the synchrotron electron densities and magnetic fields in the compact radio emitting structures are likely to be large enough to also cause sufficient synchrotron self-absorption (or `SSA') at GHz frequencies (Hodges et al.\ 1984; Mutel et al.\ 1985; O'Dea et al.\ 1991; Readhead et al.\ 1996; de Vries et al.\ 1997; Snellen et al.\ 2000; Perucho \& Mart{\'i} 2002; Artyuhk et al.\ 2008; de Vries et al.\ 2009).  Absorption by induced Compton scattering as an additional factor, in combination with FFA, has also been proposed (Kuncic et al.\ 1998).  

Despite three decades since the discovery of GPS sources as a significant population (Gopal-Krishna et al.\ 1983; Spoelstra et al.\ 1985) the issue of the nature of the absorption mechanism is still unresolved.  Stawarz et al.\ (2008), for example, comment that this is one of the important questions regarding our understanding of these sources.  Studies extending the $\nu_{\rm turnover}$ vs.\ LS anti-correlation reported by Fanti et al.\ (1990) have yielded a strong case for a self-consistent model in which CSO/GPS and CSS sources undergo self-similar growth, contain equipartition of energy, and have SSA spectral turnovers (Snellen et al.\ 2000) and statistics that conflict strongly with FFA models for the turnovers (de Vries et al.\ 2009).

Other models, on the other hand, successfully fit the $\nu_{\rm turnover}$ vs.\ LS anti-correlation with FFA turnovers by either invoking a radially decreasing ionized gas density (Bicknell et al.\ 1997) or engulfed interstellar clouds whose surfaces are photoionized by UV radiation from the lobes and the central AGN (Begelman 1999; Stawarz et al.\ 2008; Ostorero et al.\ 2010).  The linear sizes of GPS sources are comparable to that of the NELR, which contain sufficient free electrons to cause GHz FFA turnovers (van Breugel 1984), and numerous observations have revealed an association of ionized gas emission apparently coincident with the radio lobes (McCarthy et al.\ 1991; deVries et al.\ 1999; Axon et al.\ 2000; O'Dea et al.\ 2002; Jackson et al.\ 2003; Holt et al.\ 2003; Holt 2005; Morganti et al.\ 2005; Labiano et al.\ 2006; Vermeulen et al.\ 2006; Morganti 2008; Privon et al.\ 2008; Holt et al.\ 2009).  Additionally, measurements of X-ray absorption, in comparison to that of N(HI), have yielded estimates of ionization fractions of the line of sight gas on the order 90 to 99\% (Vink et al.\ 2006; Tengstrand et al.\ 2009), while significant HI absorption has been detected far more commonly in GPS galaxies than in larger radio galaxies (Pihlstr{\"o}m et al.\ 2003; Vermeulen et al.\ 2003a; Gupta et al.\ 2006; Gupta \& Saikia 2006).  The column depth of HI has also been found to be anti-correlated with LS (Pihlstr{\"o}m et al.\ 2003; Gupta et al.\ 2006; Gupta \& Saikia 2006), supporting the idea that the ambient gas is related to the $\nu_{\rm turnover}$ vs.\ LS anti-correlation.  Finally, observations of some individual GPS/CSOs have provided strong evidence in favor of FFA over SSA (e.g.\ Peck et al.\ 1999; Kameno et al.\ 2000; Kameno et al.\ 2001; Marr et al.\ 2001; Vermeulen et al.\ 2003b; Kadler et al.\ 2004; Kameno et al.\ 2005).  

\subsection{The approach in this paper}

Here, we discuss tests of both the SSA and FFA models in two CSO or CSO-candidate sources using multi-frequency VLBI observations.  Most previous attempts have focused on spectroscopic analyses, which have not been definitive because of the complexity of the composite spectra.  With our multi-frequency, high-resolution maps we have a means of separating the positional variation effects from the spectroscopic.  In particular, following the approach of Marr et al.\ (2001), through the creation of optical-depth maps at different frequencies, we are able to use the positional and spectroscopic factors independently to test the models in two ways.  

Morphologically, the optical depths in the SSA model should follow the structure in the high-frequency intensity maps.  In principle, SSA is important when the energy density in the emitted, unabsorbed synchrotron radiation would be greater than the energy density of the synchrotron electrons.  This leads to the familiar SSA turnover frequency equation in which
$$\nu_{turnover} \propto {F_\nu^{2/5} B^{1/5} \over (\theta_{maj}\theta_{min})^{2/5}}$$
(Kellermann \& Pauliny-Toth 1981), or, for a resolved source, 
$$\nu_{turnover} \propto I_\nu^{2/5} B^{1/5}.$$
And, since the intensity, $I_\nu$, is greatest where the magnetic field, B, and the particle densities are greatest SSA will cause higher frequency spectral turnovers where the raw, unabsorbed synchrotron radiation is most intense.  Therefore, the inferred SSA optical depth
at any frequency below the turnover should be largest at the positions where the high-frequency intensity is greatest.   

With FFA, though, the absorption occurs in the foreground and so the morphology in the optical-depth map will show no relation to the high-frequency intensity maps.  The free-free optical depth at any position can then be approximated, for reasonable values of temperature, as
$$\tau_{\nu,{\rm FFA}} = \int{{0.08235 ~ n_e^2 ~ T_e^{1.35} \over \nu^{2.1}} dl},\eqno(1)$$
where $\nu$ is the frequency in GHz and must be less than 10 GHz, $T_e$ is the electron temperature in Kelvin, $n_e$ is the free electron density in cm$^{-3}$, and $dl$ is distance through the source in pc (Altenhoff et al.\ 1960; Mezger \& Henderson 1967).  Although all these parameters are likely to vary throughout the source, the frequency dependence is constant and so can be pulled out of the integral, yielding
$$\tau_{\nu,{\rm FFA}} = \nu^{-2.1} \int{0.08235 ~ n_e^2 ~ T_e^{1.35} ~ dl}.\eqno(2)$$
The total effect on the spectrum along the line of sight, regardless of variation of the parameters in the integrand, should still yield an optical depth that fits $\tau_\nu \propto \nu^{-2.1}$.  
For convenience of discussion, we define the frequency-independent part of Equation 2 as $\tau_0$ so that the free-free optical depth can be rewritten as
$$\tau_{\nu,{\rm FFA}} = \tau_0~ \nu^{-2.1},\eqno(3)$$
where $\nu$ is given in GHz.  
Equation 3 separates the spatial dependence of the free-free optical depth, $\tau_0$, and the frequency dependence factor, $\nu^{-2.1}$.  With high-resolution, multi-frequency optical depths, one can then test the FFA model by determining the fit of Equation 3 to the maps.  However, we note that variations in $\tau_0$ perpendicular to line of sight on scales smaller than the synthesized beam are to be expected and will affect the apparent frequency dependence of the observed spectrum, even at VLBI resolutions.

Polarization of the radiation provides an additional test of the absorption mechanism.  If FFA dominates then there is a large column depth of free electrons which depolarize the radiation (Peck et al.\ 1999; Gugliucci et al.\ 2005).  Dual-polarization observations have, indeed, shown the detection
of polarized radiation in GPS/CSOs to be extremely rare (Peck \& Taylor 2000; Pollack et al.\ 2003; Taylor \& Peck 2003; Gugliucci et al.\ 2005, 2007; Helmboldt et al.\ 2007). 
Additionally, a number of GPS and CSS sources have been found to show large Faraday rotation measures (Kato et al.\ 1987; Stanghellini et al.\ 1998).  
With SSA, on the other hand, a net linear polarization is likely to occur with a 90$^\circ$ rotation in position angle across the spectral peak (Mutoh et al.\ 2002).  Although, if the magnetic field structure is sufficiently complex, even with SSA the total polarization could still be below detection limits.

\subsection{Previous studies of $J0029+3457$ and $J1324+4048$}

Both $J0029+3457$ and $J1324+4048$ were observed as part of the Second Caltech-Jodrell Bank VLBI survey (CJ2; Henstock et al. 1995; Taylor et al.\ 1996).
Other observations of $J0029+3457$ include visible-wavelength images (Rieke et al.\ 1979; 
Peacock et al.\ 1981; Stickel \& K{\"u}hr 1996); determination of its redshift (Zensus et al.\ 2002); determination of the GPS radio spectrum (Hutchings et al.\ 1994; O'Dea et al.\ 1991; Torniainen et al.\ 2007); VLBI images at 5 GHz (Fomalont et al.\ 2000) and at 15 GHz (Kellermann et al.\ 1998; Kovalev et al.\ 2005); 15-GHz multi-epoch VLBI images yielding no detection of growth (Kellermann et al.\ 1998, 2004); and collation of data confirming the CSO classification (Augusto et al.\ 2006).  
Other observations of $J1324+4048$ include a null detection of X-rays in the ROSAT All-Sky Survey (Britzen et al.\ 2007); a null detection of net polarization at 5.0 GHz (Pollack et al.\ 2003); a measure of the 30-GHz flux density of $0.057 \pm 0.004$ Jy, indicating a steep high-frequency spectrum with $\alpha_{5-30} \approx -1.10$ (Lowe et al.\ 2007); and apparent detection of growth of the structure between 2005 and 2008 (An et al.\ 2012).

For calculations involving the scales in the images, we assume assume H$_0$ = 70 km s$^{-1}$ Mpc$^{-1}$, $\Omega_\Lambda$ = 0.70, and $\Omega_M$ = 0.3. The luminosity distances of these sources, then, are $D_L(J0029+3457)=9.1\times10^{27}$ cm (or 3.0 Gpc) and $D_L(J1324+4048) = 8.7\times10^{27}$cm (or 2.8 Gpc), and the linear scales in the maps are 6.2 and 6.1 pc mas$^{-1}$, respectively.

\section{OBSERVATIONS AND CALIBRATIONS}

We observed $J0029+3457$ and $J1324+4048$ with the Very Long Baseline Array 
(VLBA)\footnote{The VLBA is operated by the National Radio Astronomy Observatory, which is a facility of the National Science Foundation operated under cooperative agreement by Associated Universities, Inc.} 
in combination with the Effelsberg antenna and one VLA antenna at 1.663, 2.267, 4.983, 8.417, 15.359, and 22.233, GHz on 2004 Jul 03.  The Effelsberg antenna was included only at 1.663-GHz to improve the resolution at the lowest frequency and the one VLA antenna was included for better short ({\it u,v}) spacings at all frequencies except 2.267 GHz.

Dual polarization was recorded at each station with an integration time of 2 seconds and a bandwidth of 32 MHz.  The on-source scans were spread over a wide range of hour angle to improve the ({\it u,v})-coverage.  The specifics of the observations are listed in Table 1.

Fringe fitting was performed with AIPS using 3C84 and BLLac as fringe calibrators.  We applied {\it a priori} amplitude calibration using the system temperatures and gains measured by NRAO staff.  We assume that the amplitude calibrations are accurate to within 5\%.  

For polarimetry calibration, we used 3C279 to solve for the difference in the RCP and LCP delays.  Later in the calibration we used the strong, compact source BLLac to solve for the instrumental D-terms using the AIPS task LPCAL.  

\section{MAPPING AND ANALYSIS PROCEDURES}

\subsection{Clean Maps}

Editing and mapping, using Hybrid CLEAN (Readhead \& Wilkinson 1978), were performed in DIFMAP (Shepherd et al.\ 1994, 1995; Shepherd 1997).  The data were averaged for 2 minute intervals, followed by amplitude-based flagging.  The first phase self-cal used models obtained from 1995 observations (Taylor et al.\ 1996).   
Uniform ({\it u,v}) weighting was used in the initial iterations, with small CLEAN boxes that barely encompassed obvious emission structure, and a gain of 0.05.  As the CLEAN deepened, boxes were increased as needed while excluding apparent spurious features.  Natural ({\it u,v}) weighting was used to provide deeper CLEANing when the intensity of real structure in the residual map was within 3$\sigma$.  Amplitude self-calibration was applied when the rms agreement of the model with the data was less than 1.5$\sigma.$  

For the final images, uniform weighting was used to produce higher resolution maps.  The CLEAN component models generally fit the data to within 1.0$\sigma.$  The dynamic ranges in the final maps are of order 1000:1; the largest being 3100:1 for $J0029+3457$ at 1.7 GHz and the smallest is 130 for $J0029+3457$ at 22.2 GHz.  

\subsection{Spectral-Index Maps}

Spectral-index maps were made with each pair of data sets in neighboring frequency bands.  The data were edited to yield matching ({\it u,v}) coverage and then remapped with DIFMAP.  The final CLEAN maps were made with natural weighting this time, to increase the signal to noise and decrease the uncertainties in the spectral indices.  Each pair of final CLEAN maps were convolved with identical CLEAN beams and then read into AIPS, which was used to create spectral-index maps with the task COMB.

In this way, we produced spectral-index maps between 1.7 and 2.3 GHz, 2.3 and 5.0 GHz, 5.0 and 8.4 GHz, and 8.4 and 15.4 GHz.  The 22.2-GHz maps, once the shorter spacings were cut to match the shortest 15.4-GHz spacing, were deemed to be of insufficient fidelity.  We have not, therefore, included the 22.2-GHz data in any analysis in the rest of this project.

Moderate percent errors in the CLEAN maps propagate to large errors in the spectral index, and so we masked the spectral-index maps where the signal in either CLEAN map was less than 10$\sigma$. 

Since the observations were not phase-referenced to a nearby calibrator, no absolute position information was available and so the maps could be arbitrarily shifted relative to each other.
If the maps are misaligned, extremely large and small spectral indices would result, particularly at the edges of the emission regions.  Additionally, the spectral indices across any feature will have a strong gradient.
We started with map alignments which minimized the total relative shift of the major intensity peaks.  With $J1324+4048$, this produced reasonable looking spectral-index maps with no extreme values on the edges of components and roughly symmetric $\alpha$ values.  We inferred that no need for additional shifts were needed.
With $J0029+3457$, the relative positions of the component peaks were consistent at the lower frequencies (1.7 and 2.3 GHz) and the resultant spectral index map looked reasonable.  At higher frequencies, we chose alignments that yielded the minimum magnitudes of the extreme $\alpha$'s (both positive and negative).  We also checked that our final maps did not display noticeable gradients in $\alpha$, especially at the edges of features.  

Maps of the uncertainties in the inferred spectral indices, based on both the rms in the CLEAN maps and a 5\% uncertainty in amplitude calibration, added in quadrature, and propagated through the process, were also created.  

\subsection{Optical-Depth Maps}

Data at all five frequencies were involved in creating optical depth maps at the three lower frequencies.
All data sets, for each source, were truncated at both long and short ({\it u,v}) spacings to yield the same minimum and maximum ({\it u,v}) distances and the maps were all convolved with the same CLEAN beam.  To attenuate, somewhat, the loss of resolution, the final maps were convolved with uniform weighting.  

The CLEAN maps at 15.4 and 8.4 GHz provide close approximations of the unabsorbed emission spectrum at each point in the map, and so extrapolation of these maps to lower frequencies yields reasonable maps of the unabsorbed emission at the lower frequencies.  If $\alpha_{15-8}(x,y)$ is the spectral index between 15.4 and 8.4 GHz at some position $(x,y)$ in the map, and $I_{\rm 8.4GHz}(x,y)$ is the intensity at 8.4 GHz, then the unabsorbed intensity at 2.3 GHz is 
$${\rm Unabsorbed}~I_{\rm 2.3}(x,y) = I_{\rm 8.4}(x,y)
\left[ {{\rm 2.267 GHz} \over {\rm 8.417 GHz}}\right] ^{\alpha_{15-8}(x,y)}.\eqno(4)$$  
The optical depth at 2.3 GHz at each point in the map, then, is obtained by
$$\tau_\nu(x,y)=-ln({\rm observed~} I_\nu(x,y)) + ln({\rm unabsorbed~} I_\nu(x,y)),\eqno(5)$$
and, similarly for the optical-depth maps at 1.7 and 5.0 GHz.
Optical-depth maps were made using a number of steps of the AIPS task COMB which reproduced the math of Equations 4 and 5.

The truncated data sets have much smaller ({\it u,v}) aspect ratios (ratio of longest to shortest baseline) than the original data sets 
and so this method could have resulted in the maps missing much of the structure.  Of course, with the larger beam the small scale structure was indeed not as resolved as in the full-data CLEAN maps.  One might be concerned, though, about loss of information of the large scale structure.  The detected flux densities on the remaining shortest baselines at 1.3 GHz were, in fact, 95\% of those in the full data sets.  We therefore compared the resultant CLEAN maps made with the truncated data sets with those from the full data sets and found 
no noticeable difference in the large scale structure.  

\subsection{Fits to Individual Spectra}

By stacking the maps made at identical resolutions we obtained spectra at each position in each source, which we then used to compare fits to 
simple models of FFA and SSA.  For the FFA model we fit a single free-free absorption component in the foreground of a single power-law synchrotron source, as given by
$$I_\nu = I_{2.3} \left({{\nu} \over 2.267}\right)^{\alpha}{\rm exp}(-\tau_{\rm FFA,0}~\nu^{-2.1}),\eqno(6)$$ 
where $I_{2.3}$ is the raw, unabsorbed intensity at 2.267 GHz; $\alpha$ is the raw synchrotron spectral index; $\nu$ is the frequency in GHz, and $\tau_{\rm FFA,0}$ is the FFA optical depth at 1 GHz.  For the SSA model, following Kameno et al.\ (2000), we fit our spectra to a homogeneous, single-component SSA model given by 
$$I_\nu = I_{2.3} ~ \left({\nu \over 2.267}\right)^{2.5} ~ \left[1-{\rm exp}\left(-\tau_{\rm SSA,0} ~ \nu^{\alpha-2.5}\right)\right],\eqno(7) $$
where $\alpha$ is the spectral index of the optically-thin synchrotron emission and $\tau_{\rm SSA,0}$ is the SSA optical depth at 1 GHz. We consider it highly unlikely that the structure of the synchrotron emitting region is homogeneous, but since we have only five frequencies in the spectra, a fit to a more complex SSA model is not warranted. 

\subsection{Modelfitting}

Since both sources were observed in the 1990s, we could examine our maps for movement of features over a decade.  Observations of $J1324+4048$ were made at 5 GHz in 1995 and at 15 GHz in 2001, and observations of $J0029+3457$ were made at 5 GHz in 1992 and 1994 and at 8 GHz in 1995.  Using Modelfit in DIFMAP, we fitted series of gaussian components to each complete ({\it u,v}) data set.

There are also images available on the MOJAVE database web page %(http://www.cv.nrao.edu/2cmsurvey/maps/0026+346.html, 
(Lister et al.\ 2009) which include 15-GHz VLBA observations of $J0029+3457$ from 1995 to 2002.  However, these observations are of lower resolution and smaller dynamic range and so were not useful for our analysis.

\section{RESULTS}

The results for $J1324+4048$ are clearer and easier to interpret and so, in this and following sections, we discuss the results for this source first.

\subsection {$J1324+4048$}

\subsubsection{CLEAN Maps}

The CLEAN maps at all six frequencies are displayed in Figure 1 with the peak and rms intensities listed in the figure caption.

The intensity distribution at the four lower frequencies is completely contained within two major, featureless components, which we infer are the lobes of the radio structure.  We expect that the core is located somewhere between and undetected.  The apparent structure is consistent with that found in observations by Henstock et al. (1995) and An et al.\ (2012).  At 15.4 GHz, the eastern lobe exhibits a noticeable westward elongation, although not exactly in the direction of the western lobe, and the western lobe shows a smaller elongation toward the eastern lobe.  At 22.2 GHz, both lobes appear to be resolved into major and minor components, with the minor components located towards the center, suggestive of features along the jet.  However these minor components are marginally above the noise and the reliability of this map is poor.  
Future, high sensitivity VLBI observations of this source at 22 GHz or higher could be revealing about the internal structure.

\subsubsection{Polarization}

No net polarization was detected.  The 3$\sigma$ levels in each polarization at each frequency are listed in Table 2.  These results are in agreement with those of Pollack et al. (2003) who find no detectable polarization at 5.0 GHz with a polarized intensity rms of 0.1 mJy.

\subsubsection{Spectral-Index Maps}

The spectral-index maps (where $\alpha$ is defined by $F_\nu \propto \nu^\alpha$) of $J1324+4048$ are shown in Figure 2 and the minimum and maximum uncertainties in $\alpha$ are listed in Table 3.  The uncertainty is minimum where the intensity is largest and stays close to the minimum value over most of the source.  Only near the edges of the structure do the uncertainties shoot up towards the maximum uncertainty.

The $\alpha$-map between the two highest frequencies shows that the two major components have steeply declining spectra, with $\alpha \approx -$1.0.  The spectrum in the East is slightly steeper than that in the West.  Between 8.4 and 5.0 GHz, the $\alpha$'s of both components have increased a little, but are still negative, with $\alpha \approx -$0.8.  Between 5.0 and 2.3 GHz, the spectra are essentially flat, with $\alpha \approx -$0.3.  Between 2.3 and 1.7 GHz, all $\alpha$'s are positive, indicating that the spectra of all regions have turned over.  The eastern component still has a slightly smaller $\alpha$ than the western component.

As clearly shown in Figure 2, the spectral turnover of $J1324+4048$ occurs across the entire source uniformly. This is inconsistent with the expectations for SSA, in which the brightest regions would turn over at higher frequencies.    
\subsubsection{Optical-Depth Maps}

The optical-depth maps of $J1324+4048$ are shown in Figure 3,  The uncertainties, as with the spectral-index map uncertainties, are roughly constant across most of the structure in each map and the minimum and maximum values are listed in Table 3.  
The optical depths across $J1324+4048$ are almost constant (except at the edges where the uncertainties are large).  This level of uniformity of optical depth is  suggestive of a smooth foreground of absorbing gas, and is in conflict with SSA, in which the optical depth map should reflect the 15.4-GHz intensity map.  We conclude that the optical depth maps of $J1324+4048$ provides strong evidence that FFA is significant and is the main (if not only) cause for the spectral turnover in this source.

\subsubsection{Fits to Individual Spectra}

In Figure 4, we display the spectra at the positions of the 5.0-GHz intensity peaks.  The displayed error bars correspond to the total uncertainty in the measured flux densities, due to both the 5\% amplitude calibration uncertainty and the rms in the flux densities.  Overlaid on the observed spectra are the curves representing the best reduced-$\chi$-square ($\chi_\nu^2$) fits for simple models of FFA (solid, red lines) and SSA (dashed, blue lines).  

The best-fit values of the model parameters for the spectra displayed in Figure 4 and the $\chi_\nu^2$ for these fits are listed in Table 4.  Of particular interest, the $\chi_\nu^2$'s for both locations are both significantly smaller for the FFA than the SSA model by more than a factor of two.  The $\chi_\nu^2$'s for the FFA model, considering the simplicity of the model, we take to indicate good fits since the absorbing gas most likely varies on scales smaller than the resolution in the maps, which corresponds to a linear distance of $\sim$ 15 pc.

\subsubsection{Kinematics}

We successfully fitted the data of $J1324+4048$ at 1.7 and 2.3 GHz with just two components, while at 5.0 GHz a third component was added to fit the extension of the western lobe, and a fourth component was added at 8.4 and 15.4 GHz to fit the extension of the eastern lobe.  In comparing with models from earlier observations, we considered the positions of only the two main components to be reliable.  

In Table 5 we list the separation distances between peaks and the inferred motions, using the easternmost component as the fiducial point.  The positional uncertainties listed are based on the synthesized CLEAN beam width in each direction divided by $[I_{peak}/{\rm rms}]^{1/2}$. 
We find no significant growth of the source with a conservative estimate of the maximum possible growth rate of approximately 0.03 mas yr$^{-1}$.  Ignoring the likely possibility that the growth rate was larger in the past, this implies a minimum age of 200 years.  

Multi-epoch measurements of component positions in $J1324+4048$ by An et al.\ (2012) in observations at 8.4 GHz, however, show significant growth of almost 0.1 mas between 2005 and 2009, after 12 years in which the source's size was relatively constant.  
Our inferred distance between the eastern and western components is consistent with that reported for the epochs 1995 to 2005 by An et al. The apparent growth they report is apparent only in their 2008 map, suggesting that this object has recently started a growth spurt.  Such motion would be significant for modeling the evolution of compact radio sources and so confirmation of the growth of this source with another VLBI observation could be enlightening.

\subsection {$J0029+3457$}

\subsubsection{CLEAN Maps}

The CLEAN maps at all six frequencies are displayed in Figure 5, with the peak and rms intensities listed in the caption.  The intensity distribution of $J0029+3457$ is more complex than that of $J1324+4048$.  The two lobes, located to the northeast and southwest, each have noticeable substructure and there are a couple of other features along the axis between the lobes.  Our maps are consistent with those obtained by Henstock et al. (1995), Kellermann et al.\ (1998, 2004), and Fomalont et al.\ (2000).  Within the southwest lobe, two peaks of equal intensity are apparent at 15.4 GHz but are blended at 8.4 GHz and lower frequencies, while the northeast lobe has three subfeatures apparent at 5.0 and 8.4 GHz as well as at 15.4 GHz.  One of these features appears distinct from the lobe at these frequencies, but is blended with the lobe at 2.3 and 1.7 GHz.  For the discussion in this paper we will refer to these features with the labels shown in Figure 5.d.  

\subsubsection{Polarization}

No net polarization was detected.  The 3$\sigma$ levels in our maps in each polarization at each frequency are listed in Table 2.  

\subsubsection{Spectral-Index Maps}

The spectral-index maps of $J0029+3457$ are shown in Figure 6 and the minimum and maximum uncertainties in $\alpha$ are listed in Table 3. (As with $J1324+4048$, across most of the structure the uncertainties are close to the minimum and roughly constant.) The spectra between 8.4 and 15.4 GHz at all positions except in feature `c' are declining, while that of `c' is already significantly inverted.  Feature `a' has the steepest declining spectrum, with $\alpha \approx -1.0$, while the $\alpha$'s of features `b' and `d' and the western lobe range from  $-0.5$ to $-0.1$.  The $\alpha$-map between 5.0 and 8.4 GHz is essentially the same as between the two highest frequencies, except with poorer resolution.  Between 2.3 and 5.0 GHz the spectra are less steep and feature `c' is blended with the rest of the eastern lobe and appears as an edge with a positive $\alpha$.  Between 1.7 and 2.3 GHz the spectra are essentially flat.

\subsubsection{Optical-Depth Maps}

The optical-depth maps of $J0029+3457$ are shown in Figure 7 and the minimum and maximum values of the uncertainties are listed in Table 3. 

The optical depths in $J0029+3457$ shown in Figure 7 indicate that the absorbing medium has noticeable structure.  But, this structure is unrelated to the morphology in the higher-frequency intensity maps and so this source is also inconsistent with the absorption being due to SSA.  The morphology in the optical depth maps is more easily explained as structure in the ionized component of line-of-sight gas.

\subsubsection{Fits to Individual Spectra}

In Figure 8 we display the spectra, at the positions of the 5.0-GHz intensity peaks in features `a' and `f,' overlaid with curves representing the best $\chi_\nu^2$ fits for the simple models of FFA and SSA as given by Equations 6 and 7.  The best-fit values of the model parameters are listed in Table 4.  The $\chi_\nu^2$'s are smaller for the FFA than the SSA model by approximately a factor of two.  However, we note that the extremely small $\chi_\nu^2$'s for the fits to feature `f' are not of significance since this spectrum has only a gradual bend and so fits to the spectrum are accomplished with little sensitivity to the model parameters.

\subsubsection{Kinematics}

The structure in $J0029+3457$ is relatively complex and so required up to seven components.  
The components with fitted positions that we considered reliable are located in features `a,' `b,' `c,' and two in `f,' which we label `f1' and `f2' (with `f1' closer to the center of the source).  Interestingly, component `c' was not apparent in the earlier-epoch 5.0-GHz observations and its flux density as a percentage of the other components was smaller in the 8.4-GHz observation of 1995 than in 2004, suggesting that this component brightened some time in the nine plus years before our observation.

Again using the easternmost component as the fiducial point, we list the separation distances between peaks and the inferred motions in Table 5.  
We find no significant growth in $J0029+3457$ either.  The only motion that appears greater than the uncertainty is that of `f1,' but this component is blended with `f2' and so we do not consider this apparent motion to be significant.  Our conservative estimate of the maximum possible growth rate is approximately 0.02 mas yr$^{-1}$, which is consistent with that found by Kellermann et al.\ (2004), who report an upper limit of 0.15 mas yr$^{-1}$, and yielding a minimum age estimate, ignoring the possibility of greater speeds in the past, of 1500 years. 

\section{DISCUSSION}

We have presented two tests (morphology of optical depths in comparison with intensity structure and simple model fits to individual spectra) which strongly support the model in which the spectral turnover in $J1324+4048$ is due to free-free absorption, and which are reasonably consistent with free-free absorption being the cause in $J0029+3457$.  The fits to our simple one-component model were better than expected, considering that the minor axis of the synthesized beams in our opacity maps corresponds to a linear scale of $\sim$ 15 pc.  Additionally, there is no detectable polarization. 

\subsection{$J1324+4048$:  A Strong Case for FFA}

Although the SSA model can not be definitively ruled out, keeping SSA as a significant factor in the spectral turnover of $J1324+4048$ requires a tremendous coincidence.  The structure of the synchrotron emitting region would need to have exactly the right dependence of magnetic field and electron density with radius to make the SSA optical depths appear uniform across the source as seen from our perspective, and hence
mimic that expected from FFA.  Even a model in which both SSA and FFA are important would seem contrived.

Since we have images in both the optically thick and thin spectral realms, as an additional test of the SSA model, we can compare estimates of the magnetic fields assuming equipartition, following Miley (1980), and assuming SSA, following Marscher (1983).  Using our Modelfit gaussian components at the peaks in the lobes, the equipartition magnetic field estimates, which are based on the 8.4 and 15.4 GHz data, assuming a linear depth through the source equal to the average of the two perpendicular distances, all fall in the range from 0.07 to 0.15 Gauss.  The SSA magnetic field estimation assumes that the flux density that we measure at frequencies at and below the turnover is emitted from the effective SSA photosphere at the observed frequency and hence the turnover frequency equation can be used. The modelfit parameters yield estimates ranging from 0.009 to 0.018 Gauss, an order of magnitude smaller than the B$_{equip}$ estimates.  
We find, therefore, that models that involve both SSA and energy equipartition (e.g. Snellen et al.\ 2000) result in inconsistent magnetic field estimates in $J1324+4048$.

The values of $\tau_0$ in the FFA spectral fits to the intensity peaks in $J1324+4048$ are $\approx 4$ (see Table 4).  Assuming homogeneous gas with constant electron temperature, $T_e$, in Kelvin and that the optical depth is FFA, then 
$$\tau_0 \approx 0.08235~ T_e^{-1.35} <n_e>^2 L,$$
where $<n_e>$ is the average free electron density, in units of cm$^{-3}$, and $L$ is the path length through the ionized gas in pc.  A reasonable assumption for electron temperature of 10$^4$ K, then, suggests an emission measure of free electrons of 
$$<n_e>^2 L \approx 1.2 \times 10^7 {\rm cm^{-6} pc} = 3.7\times10^{25} {\rm cm^{-5}}$$
A reasonable value of $<n_e>$ for the inner ionized regions of a radio galaxy nucleus of order $10^3$ cm$^{-3}$ requires a path length of just 12 pc.  (Measurements of ionized gas densities of $10^3$ cm$^{-3}$ and greater in GPS sources have been obtained from both X-ray absorption and optical emission line observations (e.g.\ Holt et al.\ 2003, 2011; Privon et al.\ 2008; Sawada-Satoh et al.\ 2009).)

This amount of FFA optical depth is also consistent with the ionization fractions of 90 to 99\% measured in other GPS sources (Guainazzi et al.\ 2006; Vink et al.\ 2006; Siemiginowska et al.\ 2008; Tengstrand et al.\ 2009).  An expected value of the HI column depth can be obtained from the N(HI) vs.\ LS correlation discussed by Pihlstr{\"o}m et al.\ (2003).  When their best-fit relations are applied to $J1324+4048$, which has LS $\sim$ 0.03 kpc, one obtains an expected HI column density in $J1324+4048$ of order N(HI) $\sim 4 \times 10^{20} {\rm ~cm}^{-2}$.  And, with $n_e \sim 10^3$ cm$^{-3}$ and L $\sim$ 12 pc, the free electron column depth is N$_e \sim 4 \times 10^{22} {\rm ~cm}^{-2}$, 100 times that of N(HI).   %These numbers suggest an ionization fraction of the line-of-sight gas of $\sim$ 99\%.

We therefore find that entirely reasonable physical parameters fit the observed free-free optical depths for $J1324+4048$.  

Our inferred estimate of the emission measure in combination with the observed linear size and turnover frequency can, additionally, be successfully fit to the engulfed cloud model of Begelman (1999).  According to this model, the lobes' turnover frequency of about 3 GHz (see Figure 4) and distance of $\sim$ 17 pc from the mid-point, and assuming $T_e \approx 10^4 K$, implies a UV luminosity of $L_{UV} \sim 7 \times 10^{42} {\rm ~ergs ~s}^{-1}$.  In comparison, the maximum emission measure of the FFA electrons in this model, with these parameter values, is 1 $\times 10^{-17} L_{UV} {\rm ~cm}^{-5}$, which we can set to be greater than our inferred emission measure above to yield $L_{UV} > 4 \times 10^{42}$ ergs s$^{-1}$, consistent with the inferred $L_{UV}$ using Begelman's model.

With this high emission measure, the NELR lines of H$_\alpha$, [O$_{II}$],  
and other transitions should be readily detectable, though not resolved on the same scale
as the VLBI observations presented here.  The properties of the narrow-line-region gas
have been explored in a number of CSOs (e.g., Holt et al.\ 2010; O'Dea 1998). The absorption 
of HI should in principle be detectable at these high column densities, however a problem
is that at the redshift of $J1324+4048$ (z=0.496), HI is shifted to 950 MHz which is a difficult
part of the spectrum to work in owing to emission from many fixed transmitters at this 
frequency and in the adjacent Aeronautical Mobile band.

\subsection{$J0029+3457$:  A weaker case for FFA}

The fit of the FFA model to the $J0029+3457$ maps is not as convincing.  The optical-depth maps do show structure, and so a uniform foreground of ionized gas does not work for this source.  The morphology of the optical depths is not, however, suggestive of SSA.  With SSA, we should expect the optical depths to be largest where the raw synchrotron radiation is brightest, and hence the morphology of the optical-depth maps should be somewhat similar to that of the high-frequency intensity maps, and this is clearly not the case.  These optical-depth maps, instead, seem more suggestive of foreground gas with a noticeable amount of structure.

We conclude that our observations of $J0029+3457$ supports, but does not prove, the case for free-free absorption as the main cause of the spectral turnover.

The values of $\tau_0$ in the FFA spectral fits to the intensity peaks of $J0029+3457$ are 0.9 and 1.4, less than half that inferred for $J1324+4048$ and so this optical depth is easily accomplished with a path length of order 5 pc.

The linear extent of $J0029+3457$ is about 0.18 kpc, which, according to the best-fit N(HI)--LS relation by Pihlstr{\"o}m et al.\ (2003),
implies an expected N(HI) for this source of order 2 $\times 10^{20} {\rm ~cm}^{-2}$.
And, with $n_e \sim 10^3$ cm$^{-3}$ and L $\sim$ 5 pc, the free electron column depth is N$_e \sim 1.5 \times 10^{22} {\rm ~cm}^{-2}$, suggesting an ionization fraction of 98\%, also consistent with observations of other GPS sources (Guainazzi et al.\ 2006; Vink et al.\ 2006; Siemiginowska et al.\ 2008; Tengstrand et al.\ 2009).

In applying Begelman's engulfed cloud model to $J0029+3457$, we must consider the two lobes separately since they are different distances from the apparent core component.  We again find our measurements to be consistent with this model.  The linear distances from the core and the turnover frequencies of the easternmost and westernmost components are $R \approx 0.04$ and $0.19$ kpc and $\nu_{\rm turnover} \approx 2$ and $3$ GHz, respectively.  Assuming $T_e \approx 10^4 K$, applying these parameters to this model yields expected values of the AGN UV luminosity of $L_{UV} \sim 2 \times 10^{43}$ and $9 \times 10^{44} {\rm ~ergs ~s}^{-1}$.  The inferred values of $\tau_{\rm FFA,0} \approx$ 1.4 and 0.9 imply emission measures of $\sim 1.3 \times 10^{25}$ and $8.5 \times 10^{24} {\rm cm}^{-5}$, which, by Begelman's model, suggests lower limits to the UV luminosity of $\sim 7 \times 10^{42}$ and $1 \times 10^{44} {\rm ~ergs ~s}^{-1}$, less than the expected $L_{UV}$ above.

The redshift of $J0029+3457$ (z=0.517) is about the same as that of $J1324+4048$ (0.496) and so detection of HI poses the same observational problem.  The inferred emission measure in $J0029+3457$ is about half of that in $J1324+4048$, which is still large enough that the NELR lines should be detectable.

\subsection{The Self-Absorbed Core Component of $J0029+3457$}

Unlike the rest of the source, component `c' in $J0029+3457$ has a spectrum which is already inverted between 8.4 and 15.4 GHz.  Typically, a compact component between the lobes with an inverted spectrum at high frequencies is inferred to indicate the location of a self-absorbed core.  The cause for the inverted spectrum of `c' does indeed seem unrelated to the spectral turnover of the rest of the source. 

If this is the core, one might expect its location to be closer to the midway point between the lobes.  Considerable range in arm-length ratios, though, have been found in CSOs (Readhead et al.\ 1996, Tremblay et al.\ 2010).  These are most likely the result of inhomogeneity in the ambient medium.  Applying the model of Swarup and Banhatti (1981), in which asymmetric jet lengths result from variations in jet power and ambient gas density, to our modelfit values at 8.4 and 15 GHz of components 'a' and 'f2' of the flux densities and distances from component `c,' we obtain $L_{SW}/L_{NE} \sim 2.6$ and $\rho_{NE}/\rho_{SW} \sim 0.015$, where L is the jet power and $\rho$ is the ambient gas density in each direction.

Our modelfit parameters also allow for a magnetic field estimate in component `c', assuming it to be self-absorbed, following Marscher (1983).  Since the minor axis is unresolved, even at 15.4 GHz, we obtain an upper limit of 3 Gauss along a major axis of linear length $\approx$ 2 pc.

\subsection{Linear Motions}

Our inferred maximum growth rate of each source, which we assume is twice that of the motion of the end components relative to the core, suggest maximum linear speeds of 0.3c and 0.2c in $J1324+4048$ and $J0029+3457$, respectively.  These upper limits are not significant since they are greater than those typically measured of hot spots at the end of jets in CSOs, which are generally around 0.1c and have been measured as low as 0.024c (Taylor et al.\ 2009).

\section{CONCLUSIONS}

By extrapolating the spectra in our two highest frequency maps to low frequencies and comparing to our observed maps at these low frequencies, we obtained optical-depth maps of $J1324+4048$ and $J0029+3457$ at three frequencies.  In $J1324+4048$ these maps consistently show uniform optical depths across the source and the frequency dependence agrees with that of free-free absorption.  These maps, therefore, fit a model of free-free absorption by a smooth foreground 
of ionized gas.  Additionally, no net polarization is detected anywhere in the maps, consistent with a significant foreground of free electrons.  Reasonable values for temperature (10$^4$ K), electron density ($\sim$ 10$^3$ cm$^{-3}$), and path-length ($\sim$ 12 pc) fit the observed optical depths.  We conclude that the case for free-free absorption as the
cause for the spectral turnover in $J1324+4048$ is very strong. 

The optical-depth maps of $J0029+3457$ show some structure, but this structure is unrelated to that displayed in the higher-frequency intensity maps, suggesting that the absorption of the lower frequency radiation is not SSA.  No net polarization is detected in $J0029+3457$ either.  Considering the obvious structure in $J0029+3457$, FFA is considered a reasonable possibility with variation in column depths on scales smaller than our synthesized beam.

There is a component near the eastern lobe in $J0029+3457$, with a spectrum which is inverted even at our highest frequencies, seemingly unrelated to the spectral turnover of the rest of the source.  We propose that this component is likely a self-absorbed core, even though this component is not located near the mid-point between the two ends of the source.  Assuming SSA in the core, we obtain an upper limit to the magnetic field of 3 Gauss at a radius of order 1 pc.

No detectable motion is apparent in either source, with upper limits of 0.03 and 0.02 mas yr$^{-1}$, corresponding to maximum linear speeds of 0.3c and 0.2c and yielding age limits of 200 and 1500 years for $J1324+4048$ and $J0029+3457$, respectively.

\acknowledgements
This research was aided, in part, by funding from the Research Corporation.  We are grateful for the friendly assistance of the staff at the National Radio Astronomy Observatory and that provided by Michael Gillin.  
%The data obtained for this paper was obtained with observations made possible by the National Radio Astronomy Observatory, which is a facility of the National Science Foundation operated under cooperative agreement by Associated Universities, Inc. 

\clearpage

%Figure1
\begin{figure}
\phantom{mark}
\vspace{17.5cm}
\includegraphics{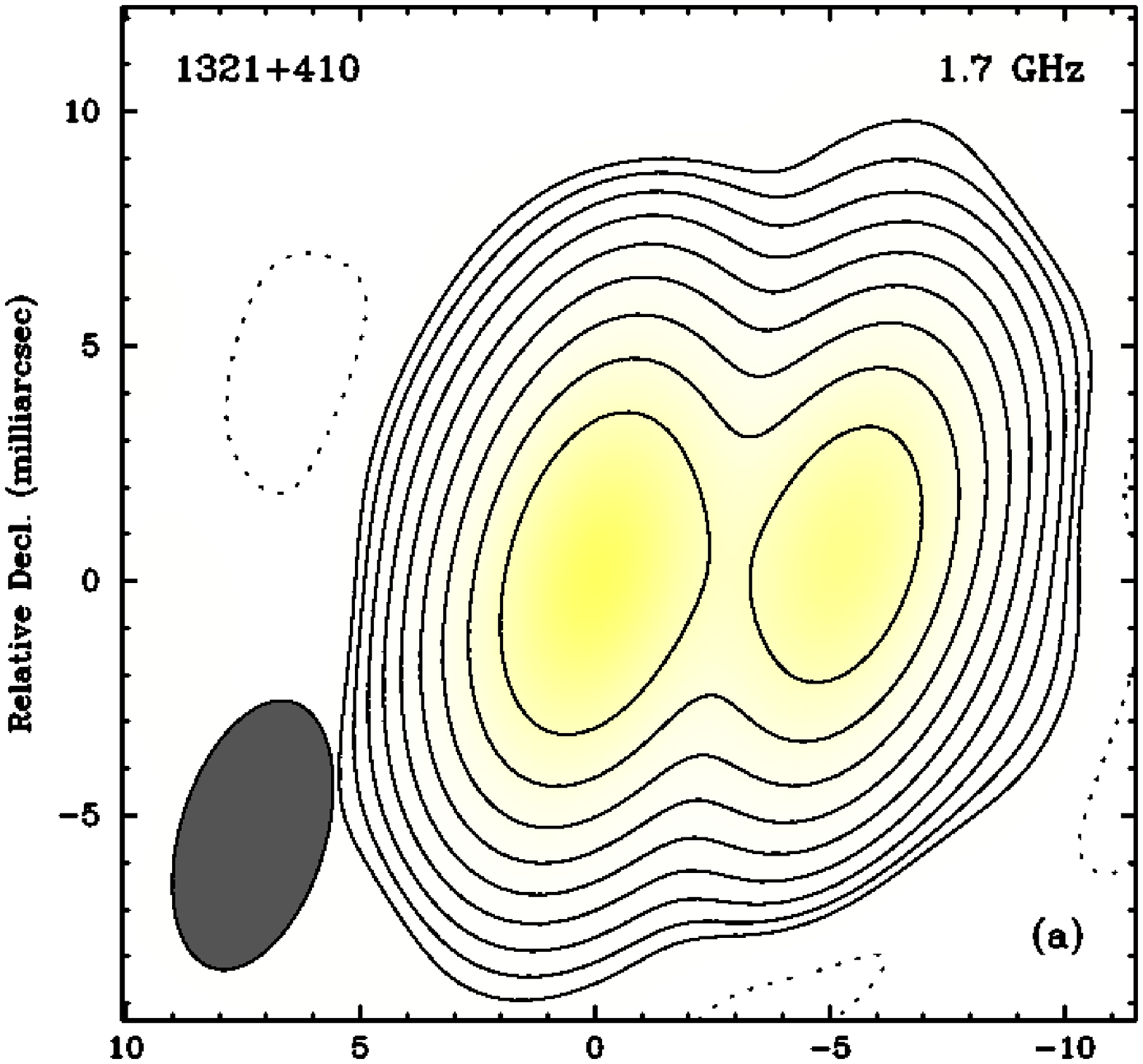}
\includegraphics{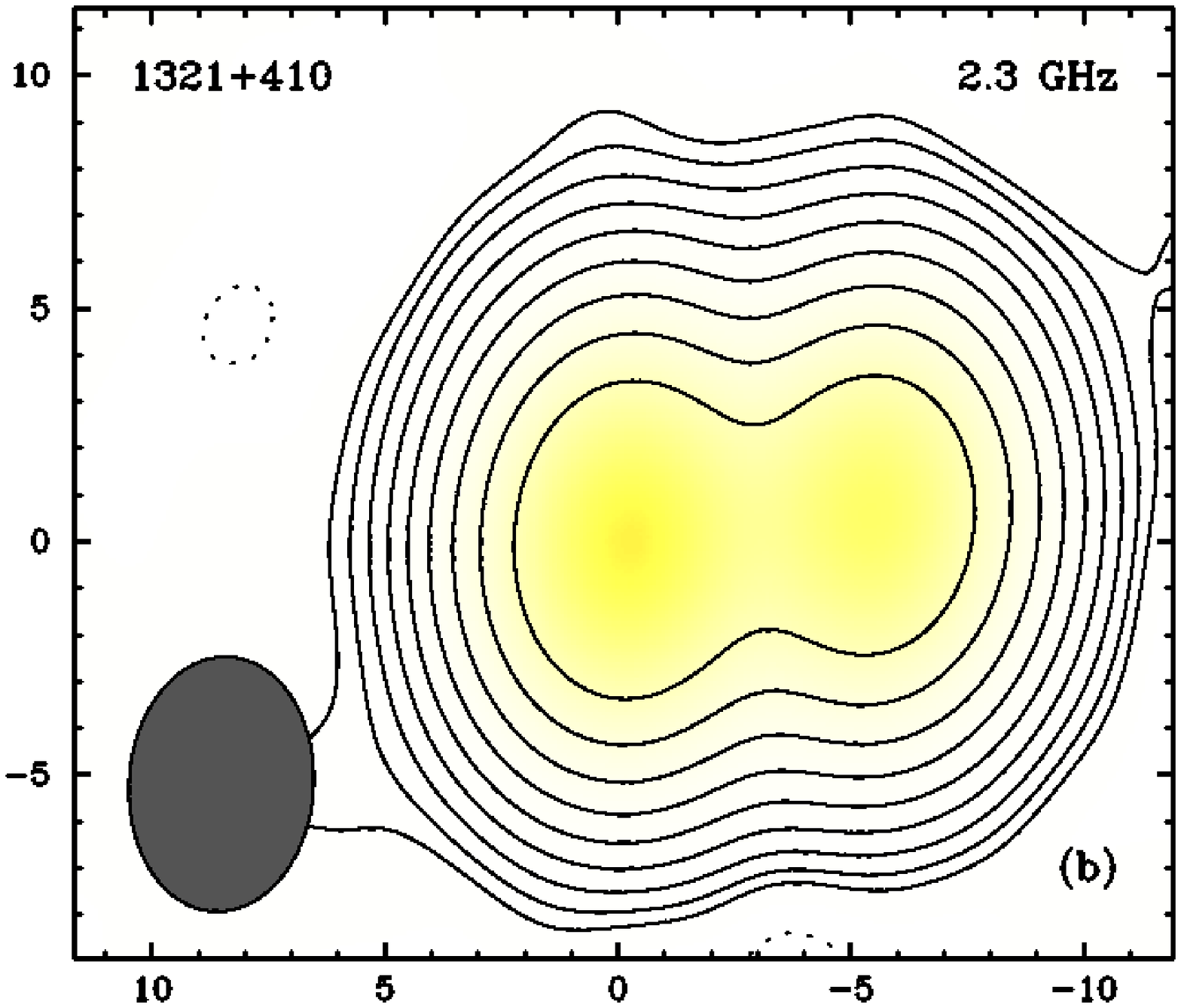}
\includegraphics{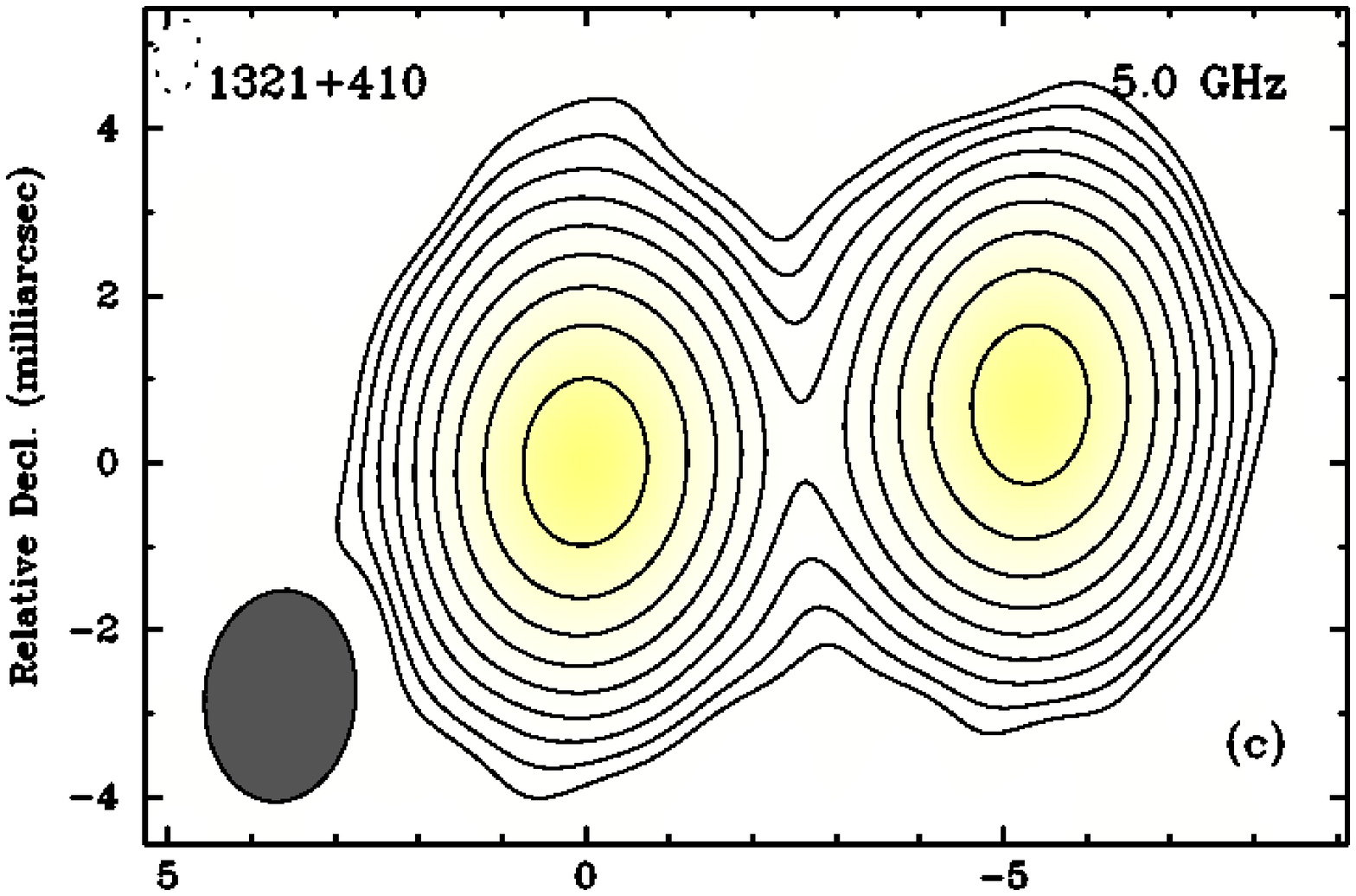}
\includegraphics{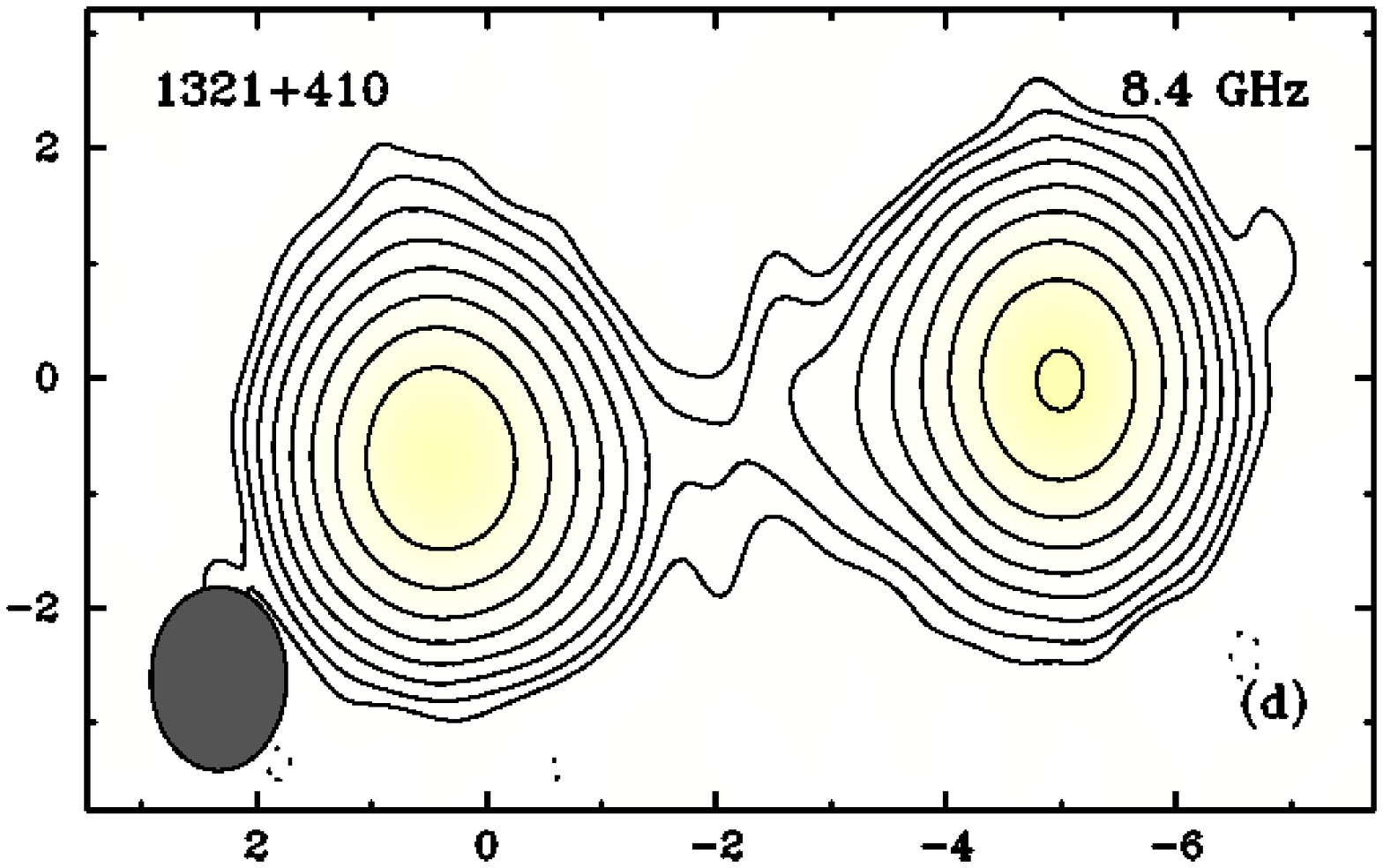}
\includegraphics{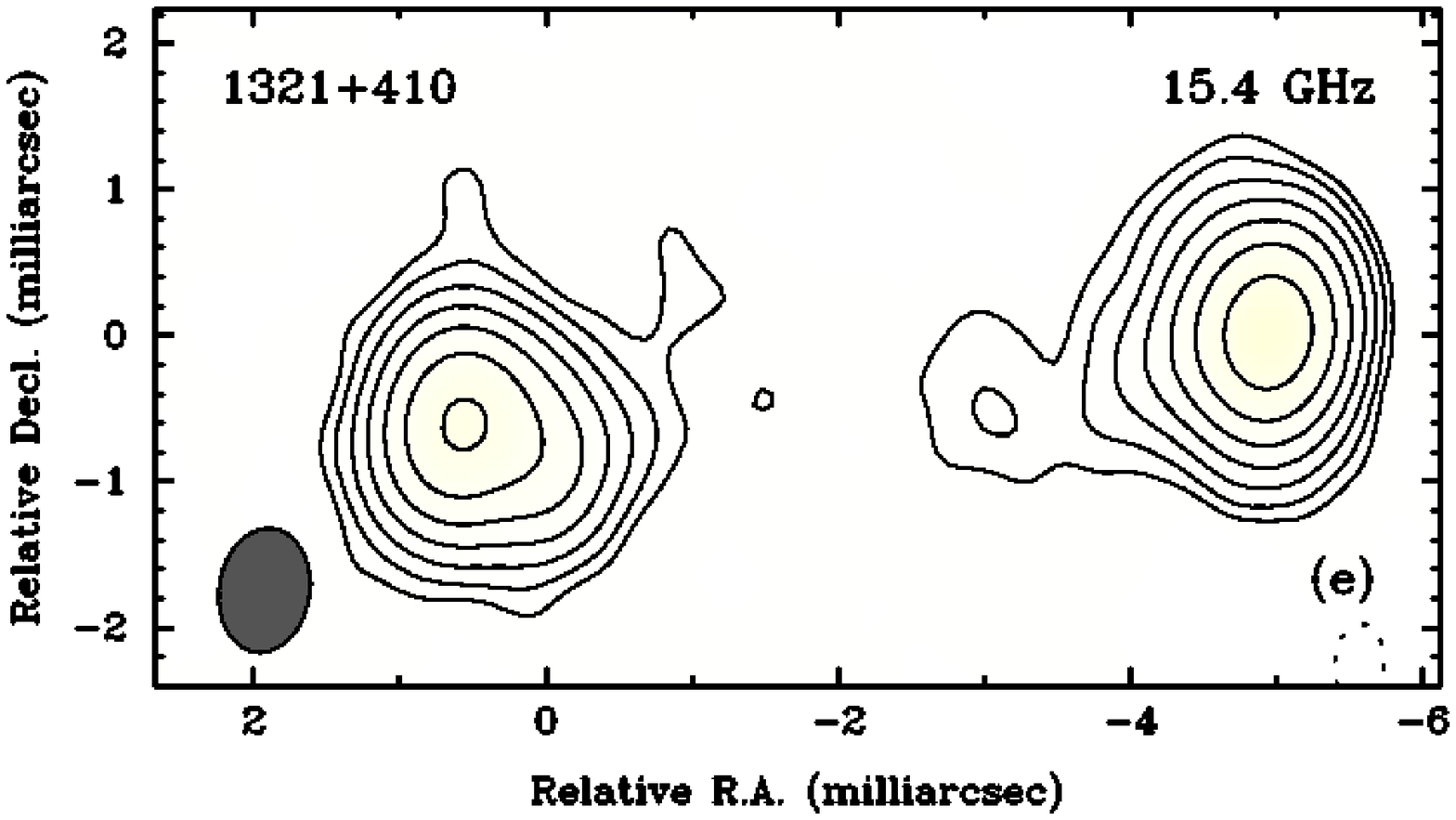}
\includegraphics{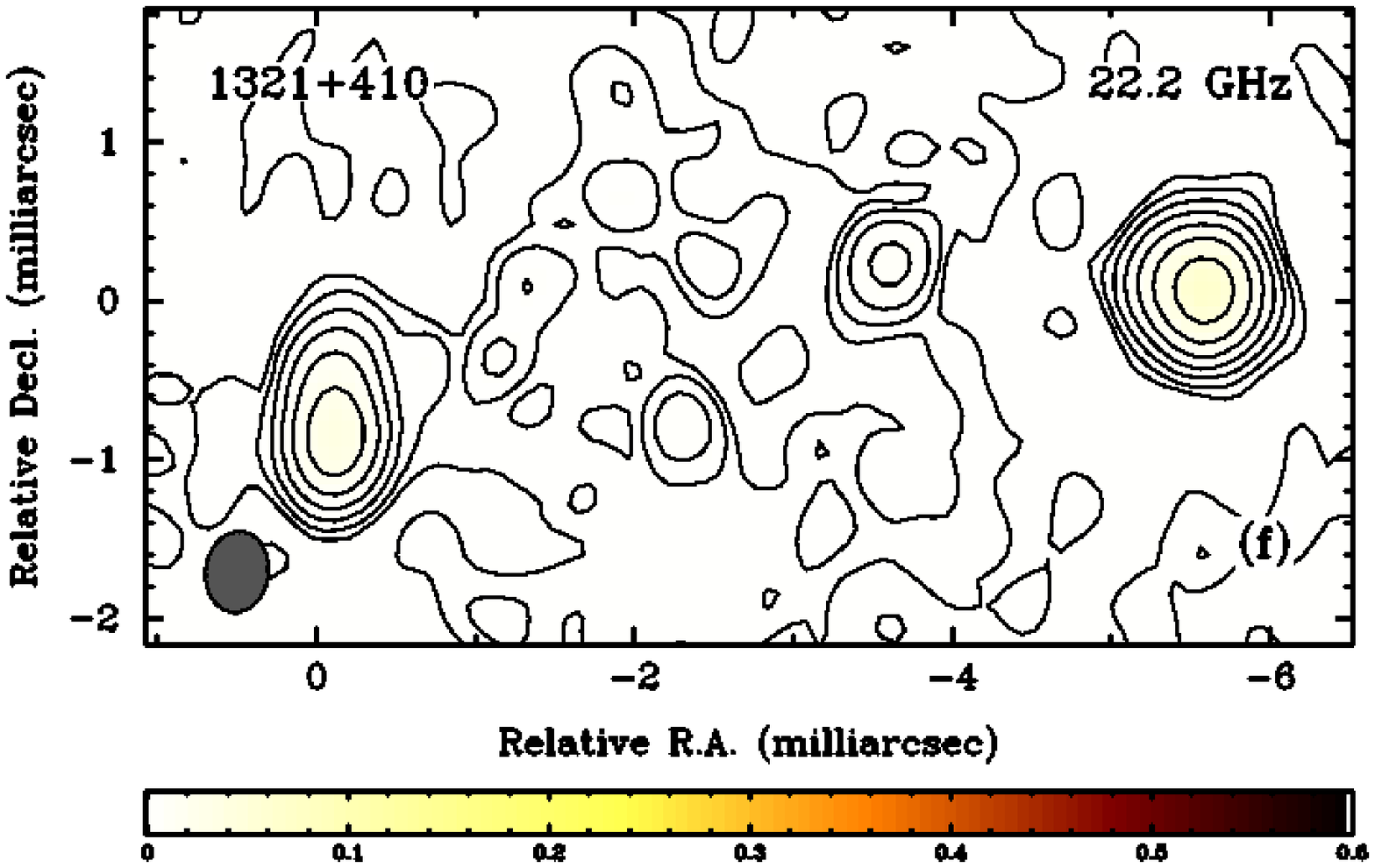}
\caption{CLEAN intensity maps of $J1324+4048$ at all frequencies, made with uniform weighting:
(a) 1.7 GHz; (b) 2.3 GHz; (c) 5.0 GHz; (d) 8.4 GHz; (e) 15.4 GHz; and (f) 22.2 GHz.
The displayed contours represent $-$5, $-$2.5, 2.5, 5,
10, 20, 40,...  times the rms.  
The peak values and rms noise levels in each map are:
(a) 0.209 \& 0.00013 Jy per beam; (b) 0.251 \& 0.00014 Jy per beam;
(c) 0.174 \& 0.00018 Jy per beam; (d) 0.103 \& 0.00015 Jy per beam;
(e) 0.047 \& 0.00019 Jy per beam; and (f) 0.0666 \& 0.00027 Jy per beam.
The FWHM of the CLEAN beams (displayed in the lower left corners)
are:
(a) 5.91$\times$3.09 mas, @ 19.96$^\circ$; 
(b) 5.47$\times$3.96 mas, @ $-$3.78$^\circ$;
(c) 2.53$\times$1.80 mas, @ $-$4.76$^\circ$; 
(d) 1.60$\times$1.18 mas, @ $-$0.25$^\circ$;
(e) 0.85$\times$0.62 mas, @ $-$8.43$^\circ$; 
(f) 0.51$\times$0.40 mas, @ $-$7.70$^\circ$.
The maps are not all displayed to the same scale.}
\end{figure}

%Figure2
\begin{figure}
\phantom{mark}
\vspace{10.0cm}
\includegraphics{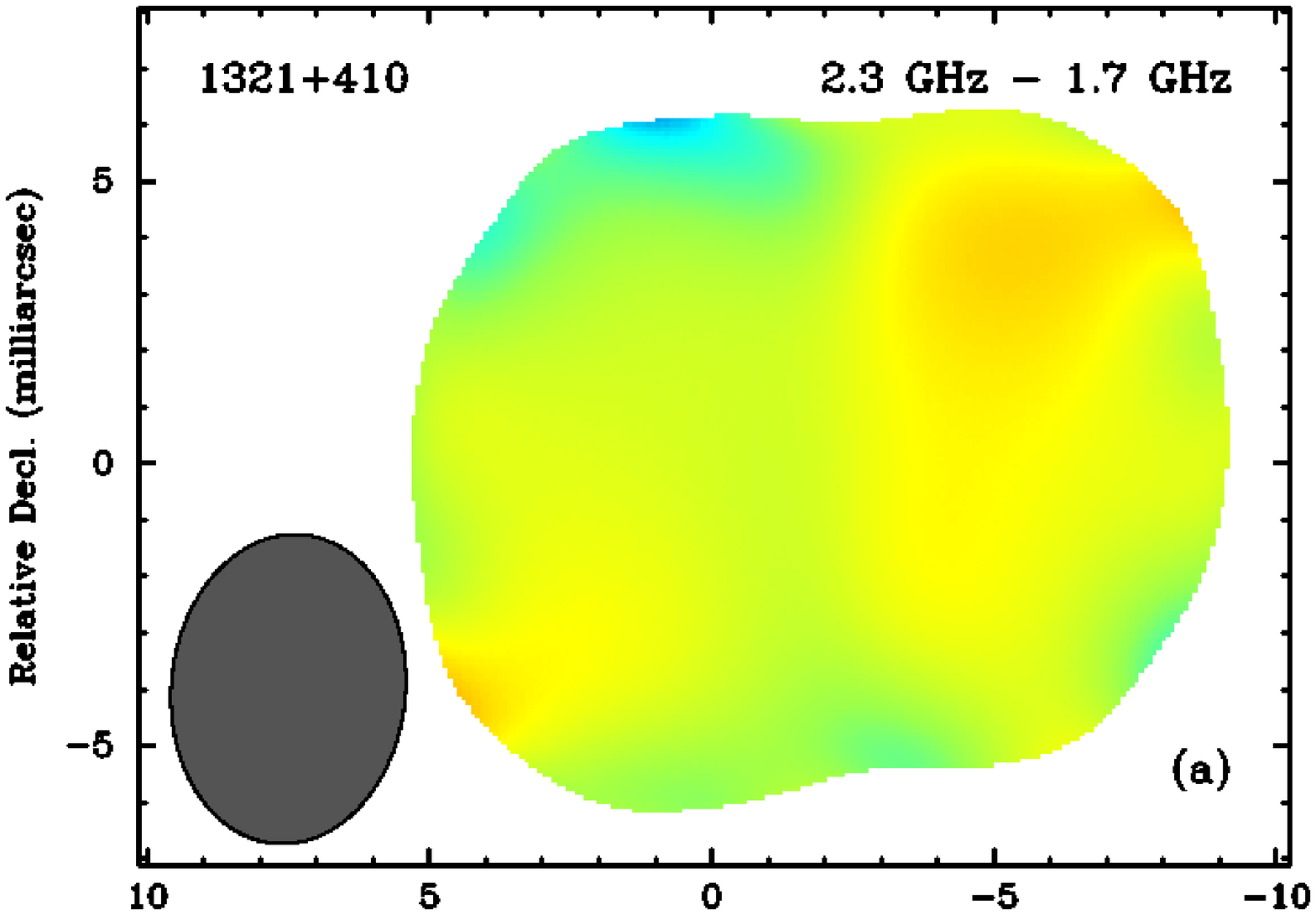}
\includegraphics{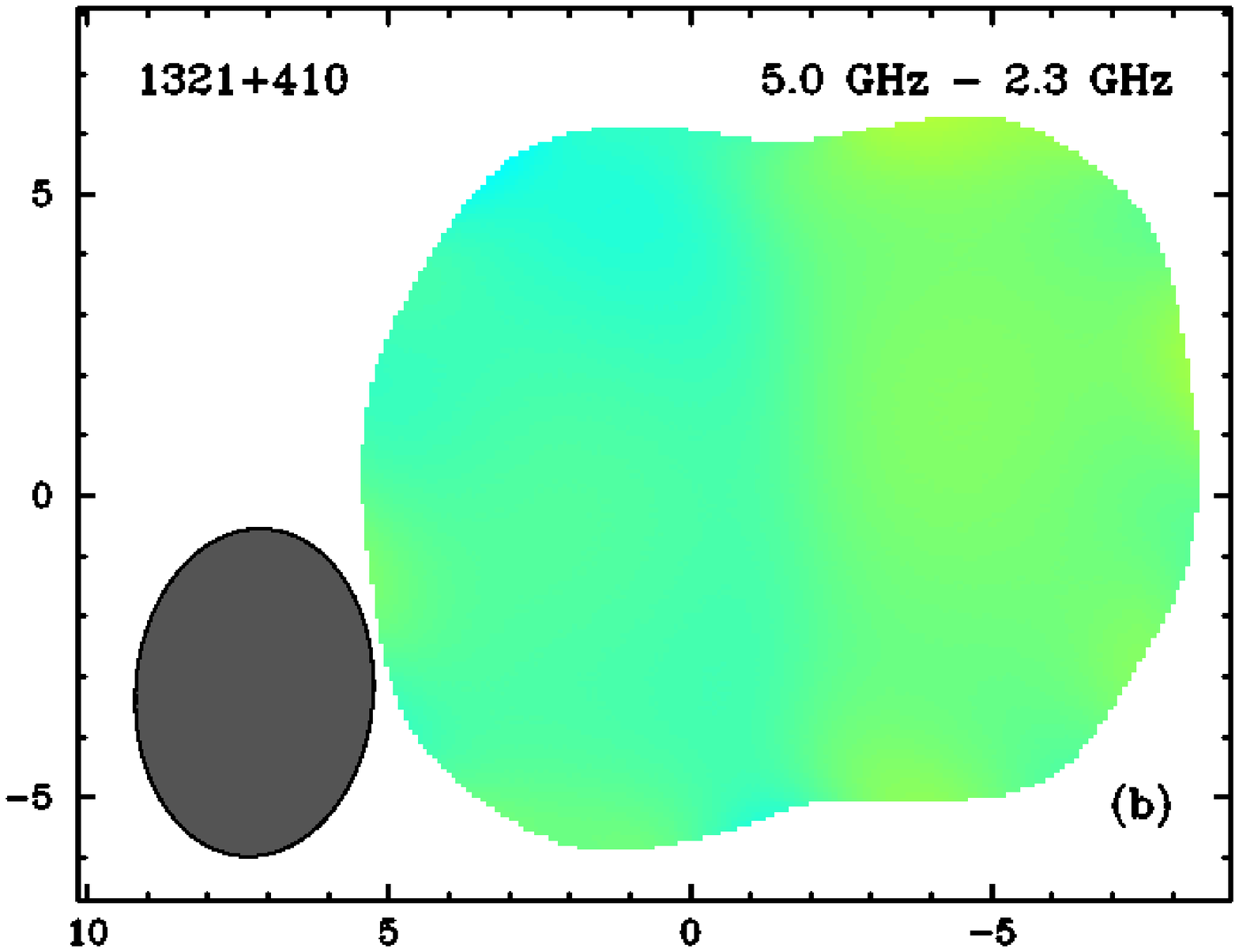}
\includegraphics{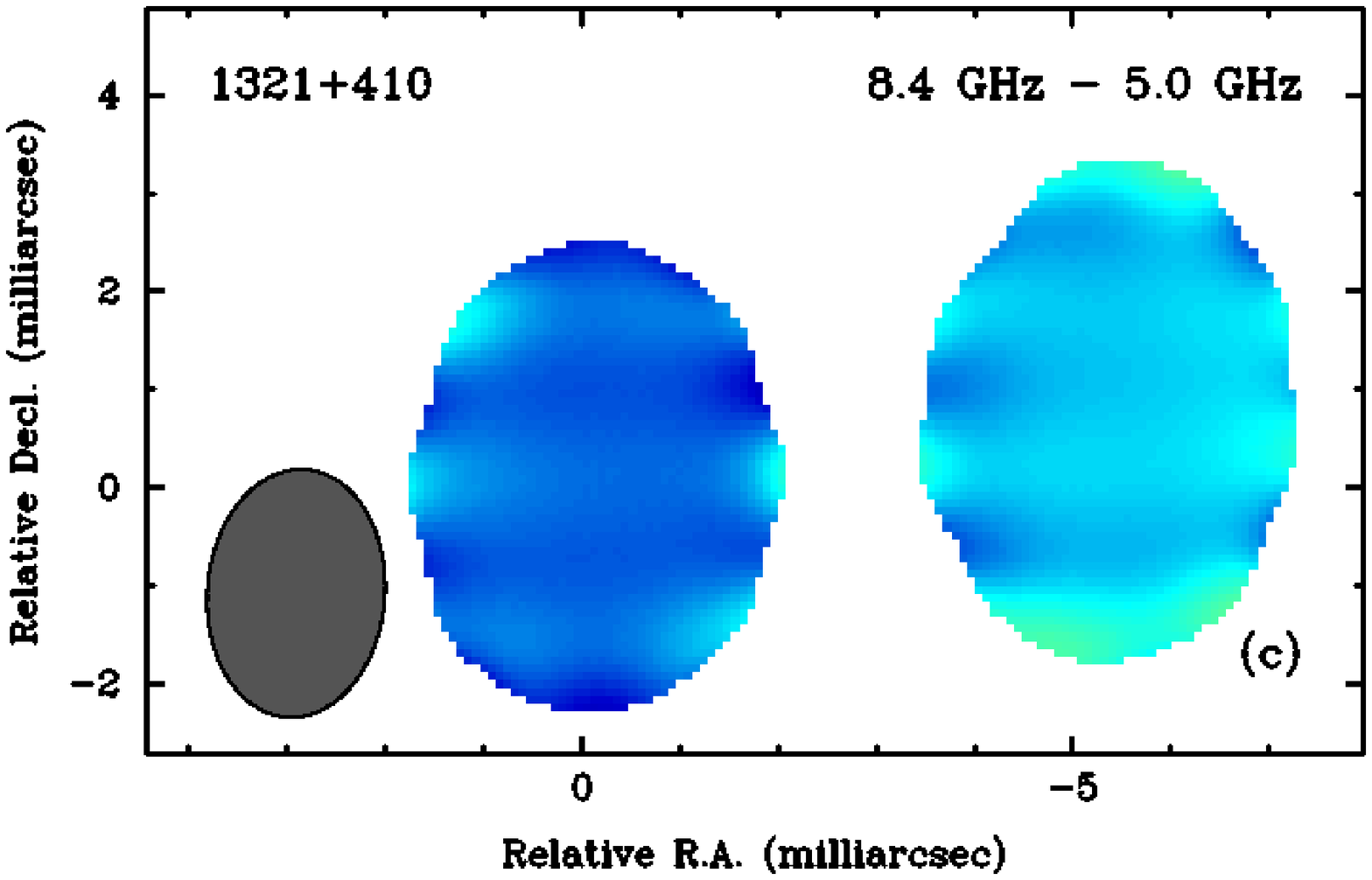}
\includegraphics{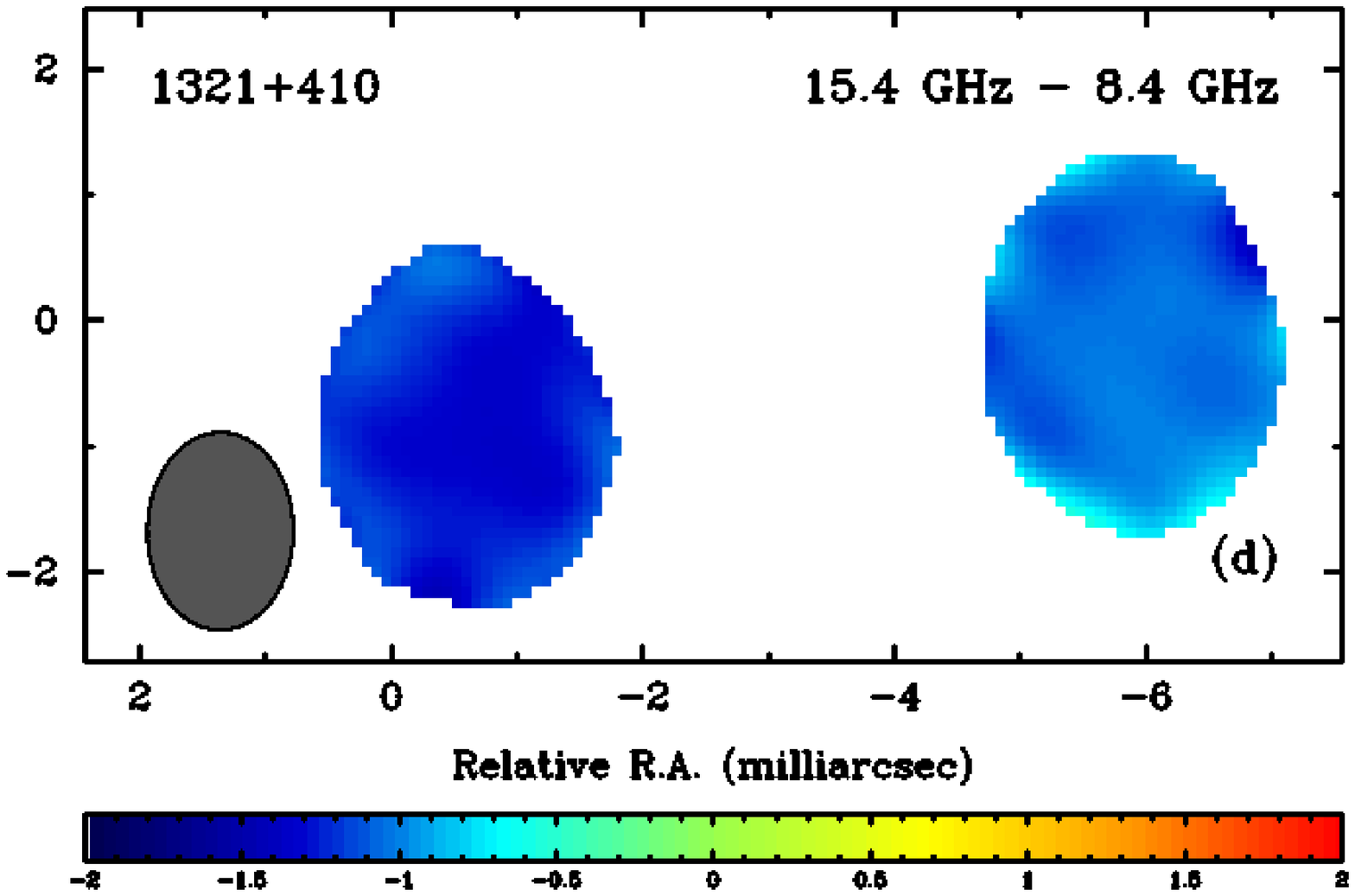}
\caption{Spectral-index maps of $J1324+4048$:
(a) between 2.3 and 1.7 GHz; (b) between 5.0 and 2.3 GHz;
(c) between 8.4 and 5.0 GHz; and 
(d) between 15.4 and 8.4 GHz.  
The colors indicate spectral index values, $\alpha$, as indicated on the wedge where
F$_\nu \propto \nu^\alpha$.  The bluer colors indicate more negative spectral indices (i.e. steeper, optically thin spectra).  The uncertainties in the spectral indices are listed in Table 3.  The input CLEAN maps were both convolved with CLEAN beams displayed
in the lower left corners.  The maps are not all displayed to the same scale.}
\end{figure}

%Figure3
\begin{figure}
\vspace{6cm}
\includegraphics{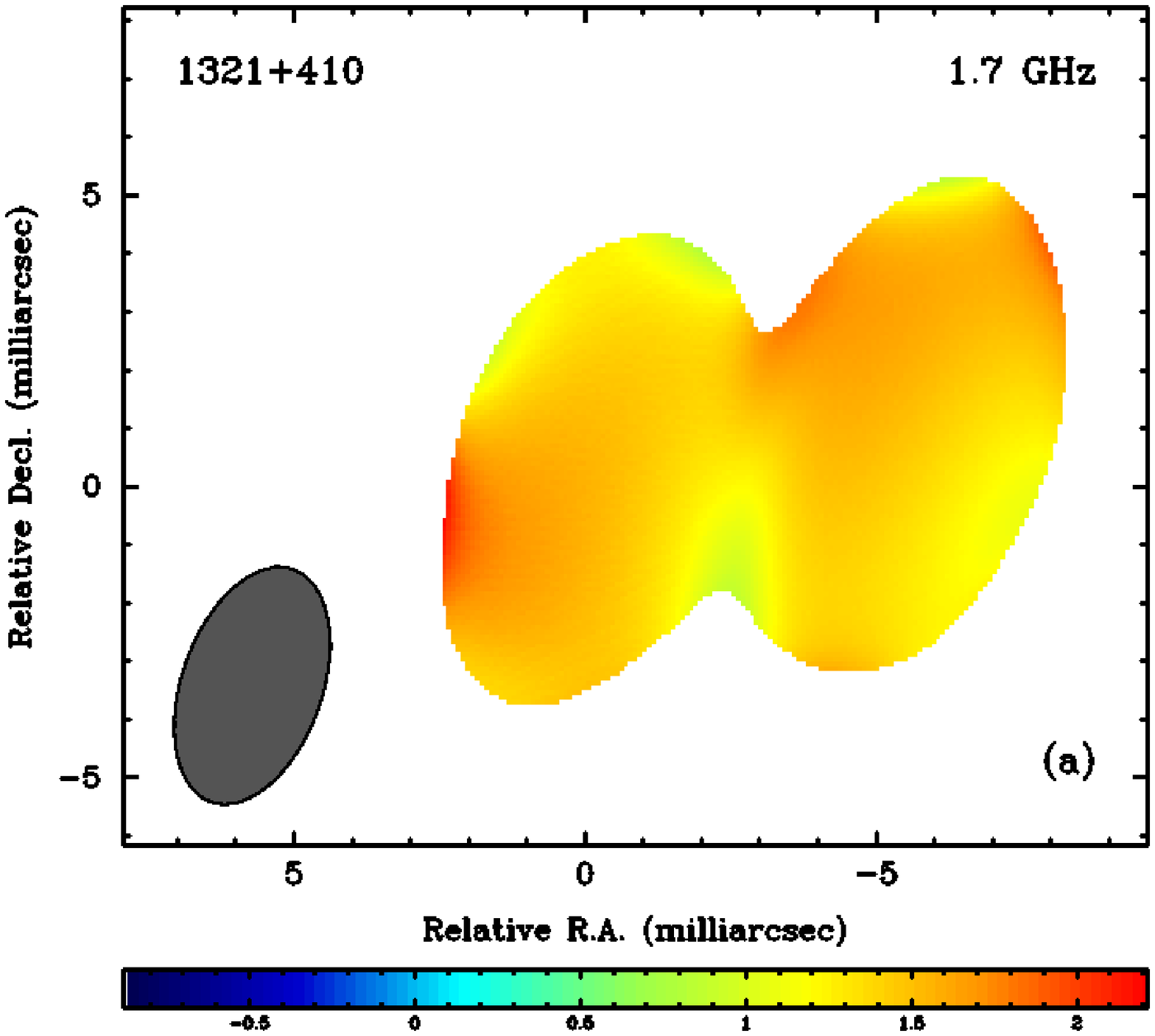}
\includegraphics{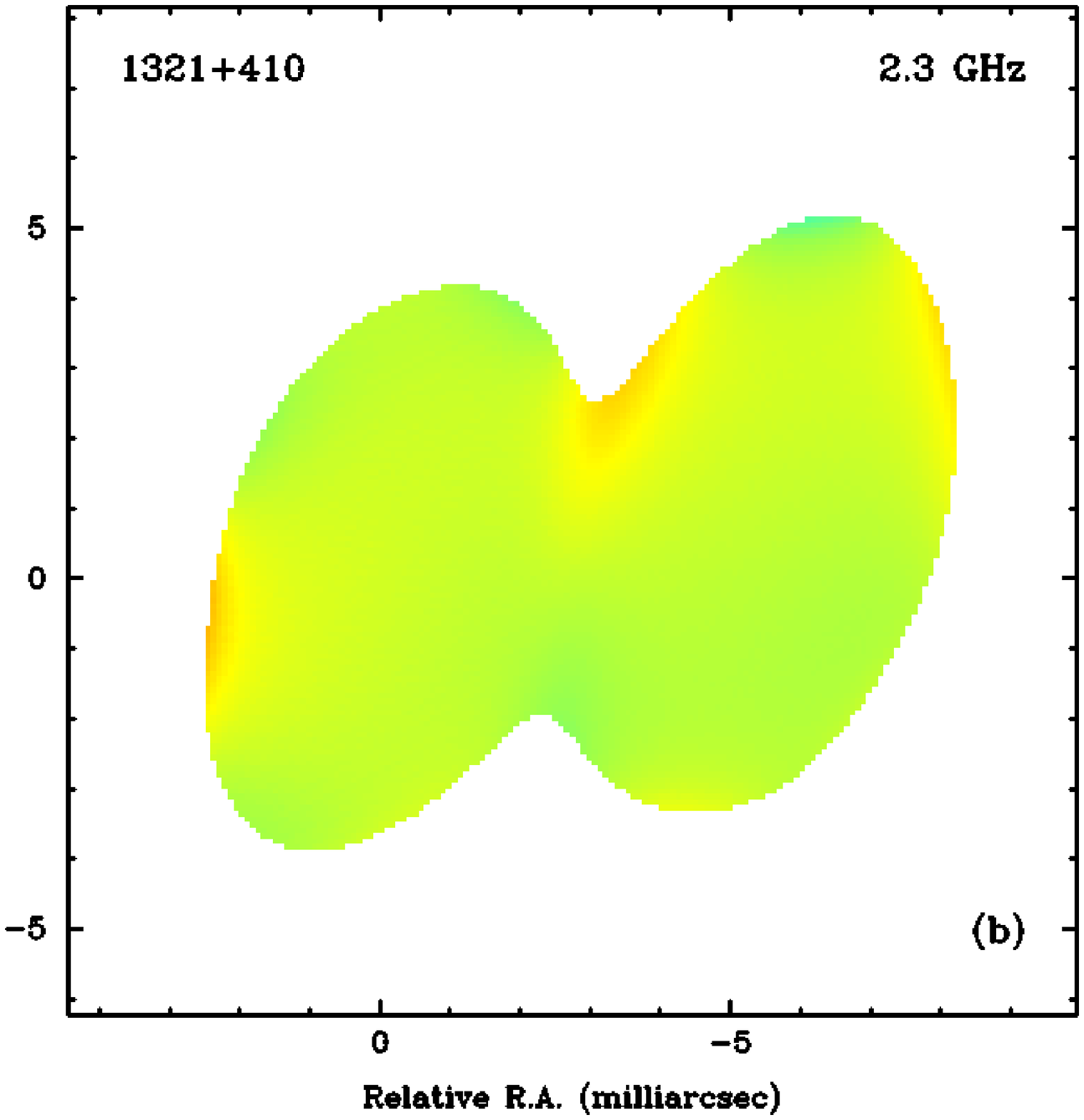}
\includegraphics{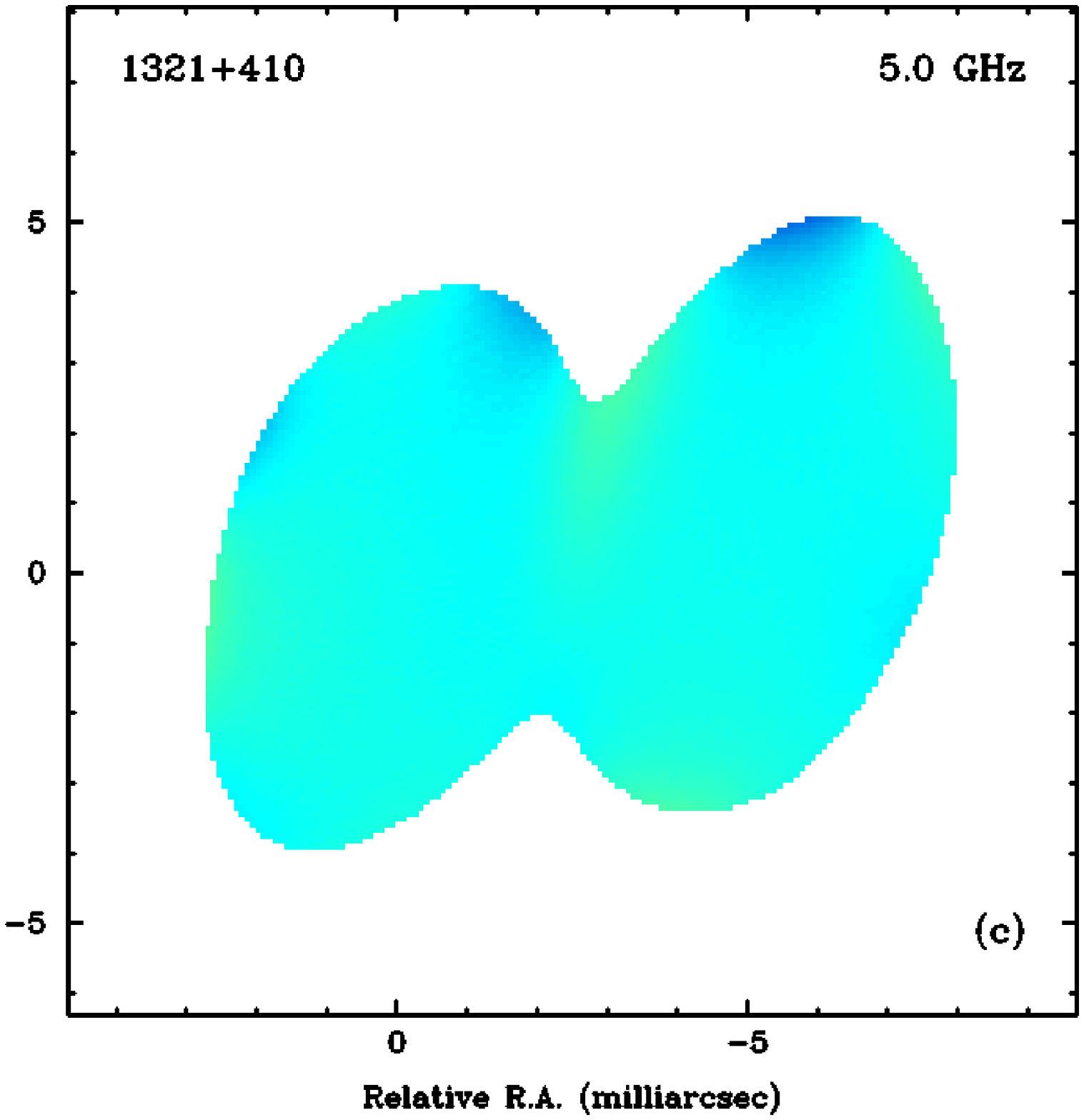}
\caption{Optical-depth maps of $J1324+4048$ at (a) 1.7 GHz; (b) 2.3 GHz; and (c) 5.0 GHz.  The colors indicate optical depth values as indicated on the color wedge with redder colors indicating larger optical depths.  The uncertainties in the optical depths are listed in Table 3.}
\end{figure}

%Figure4
\begin{figure}
\vspace{6cm}
\includegraphics{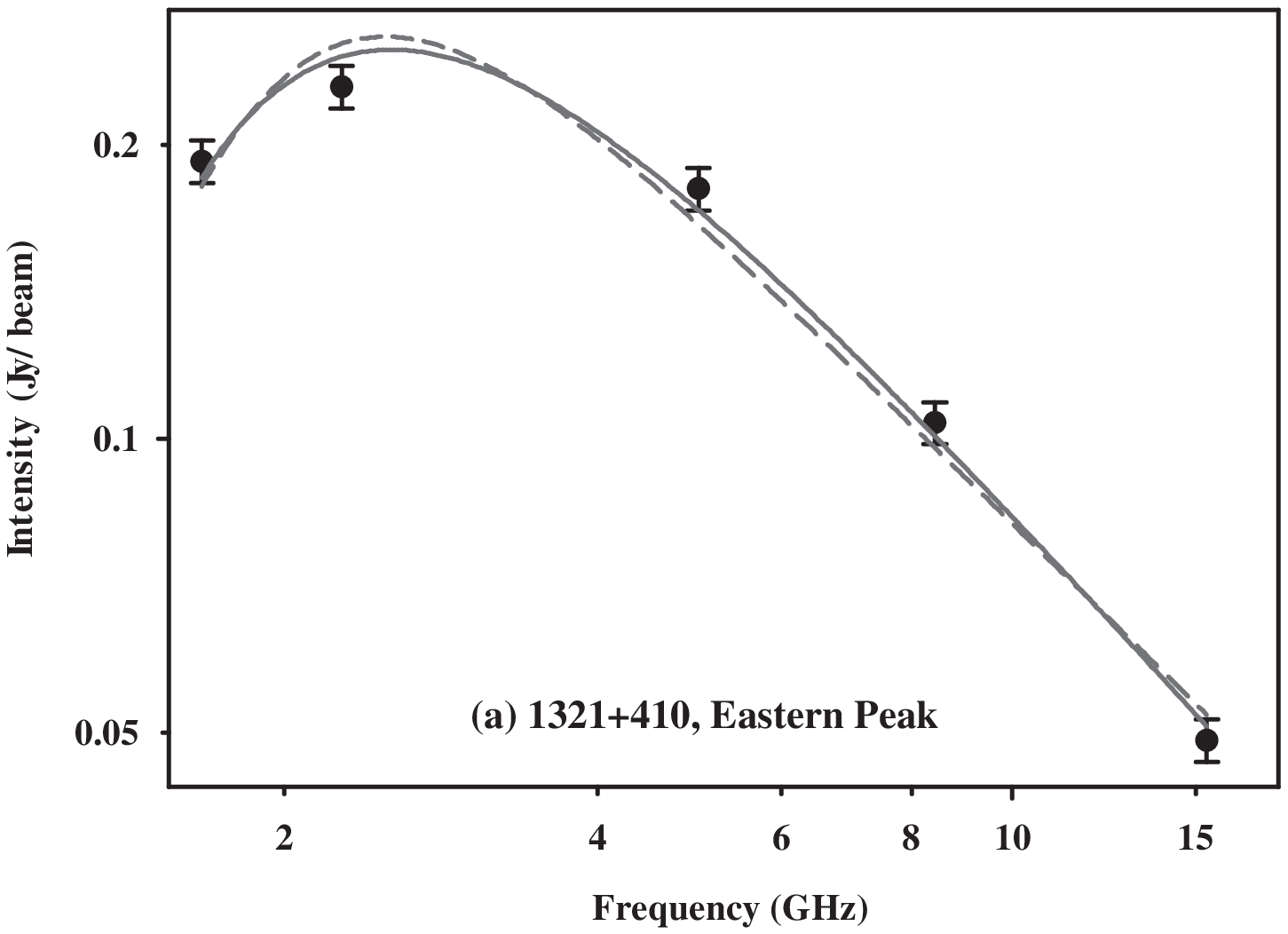}
\includegraphics{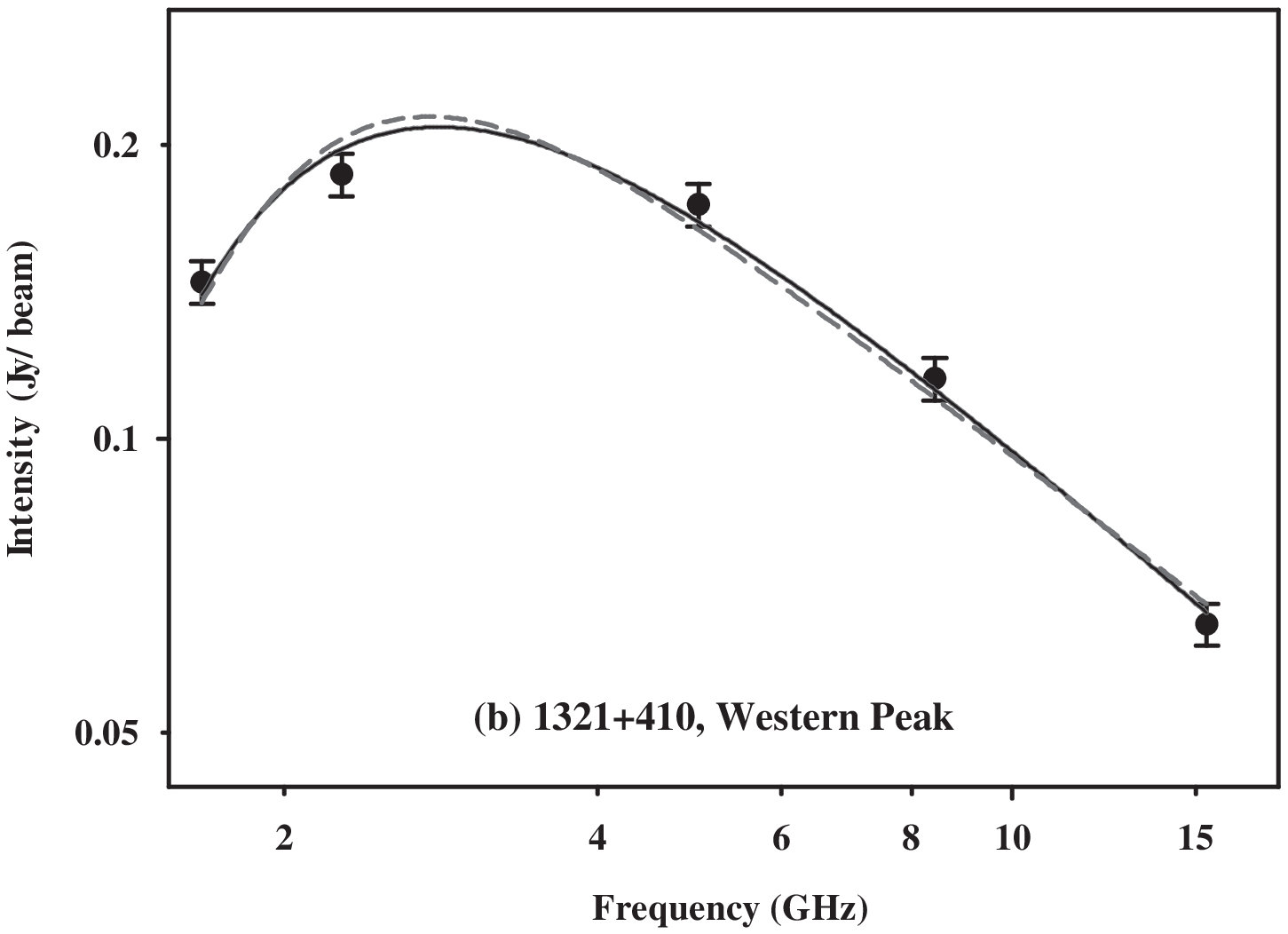}
\caption{Log-log plots of the spectra at the positions of the (a) eastern peak and (b) western peak in emission in $J1324+4048$ at 5.0 GHz.  The error bars represent the total intensity uncertainties, including
a 5\% uncertainty in the amplitude calibration.  The overlaid smooth lines are the curves of the best-fit models of a single FFA component (solid blue) and of a single SSA component (dashed red).}
\end{figure}

%Figure5
\begin{figure}
\phantom{mark}
\vspace{17.5cm}
\includegraphics{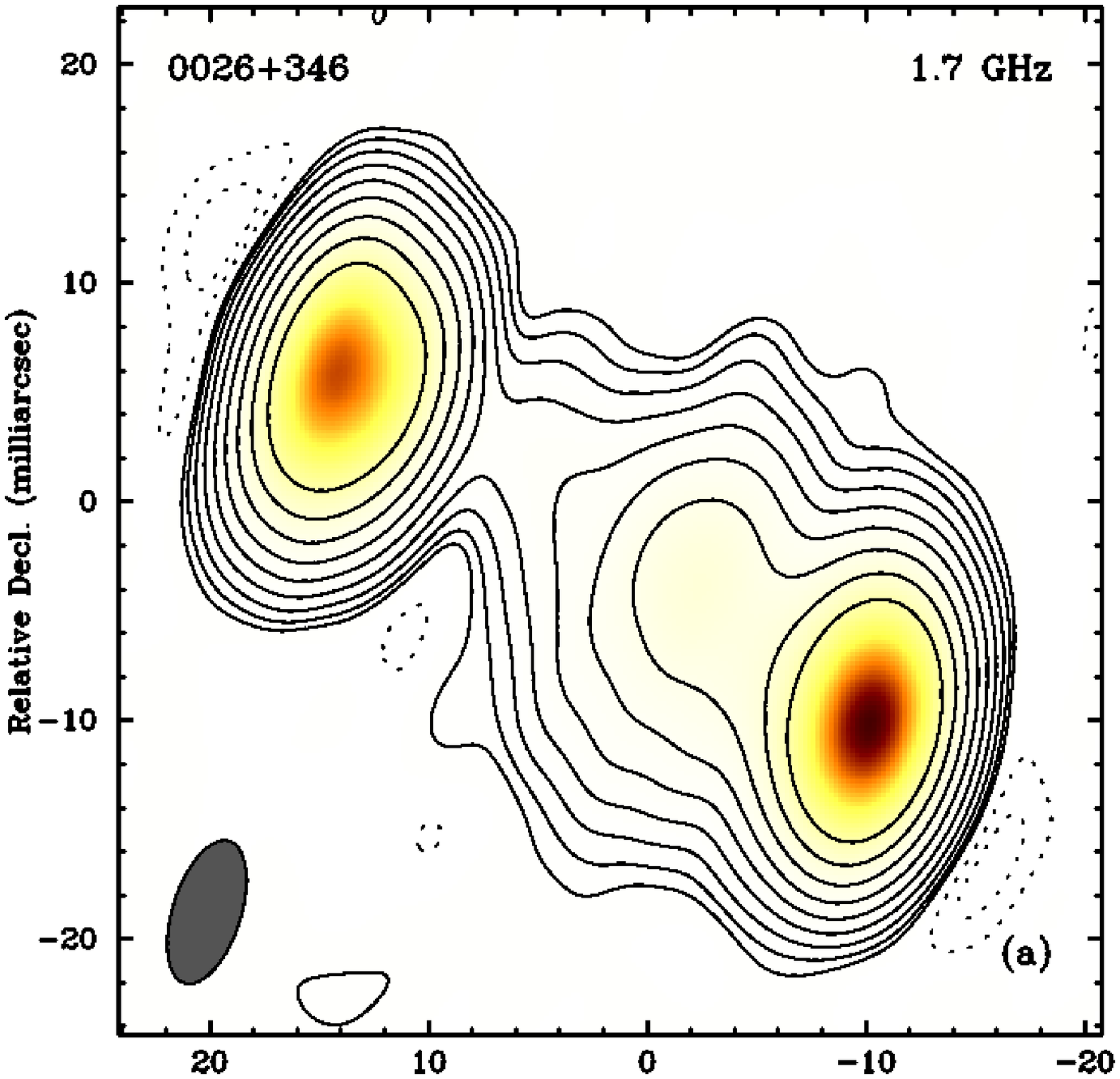}
\includegraphics{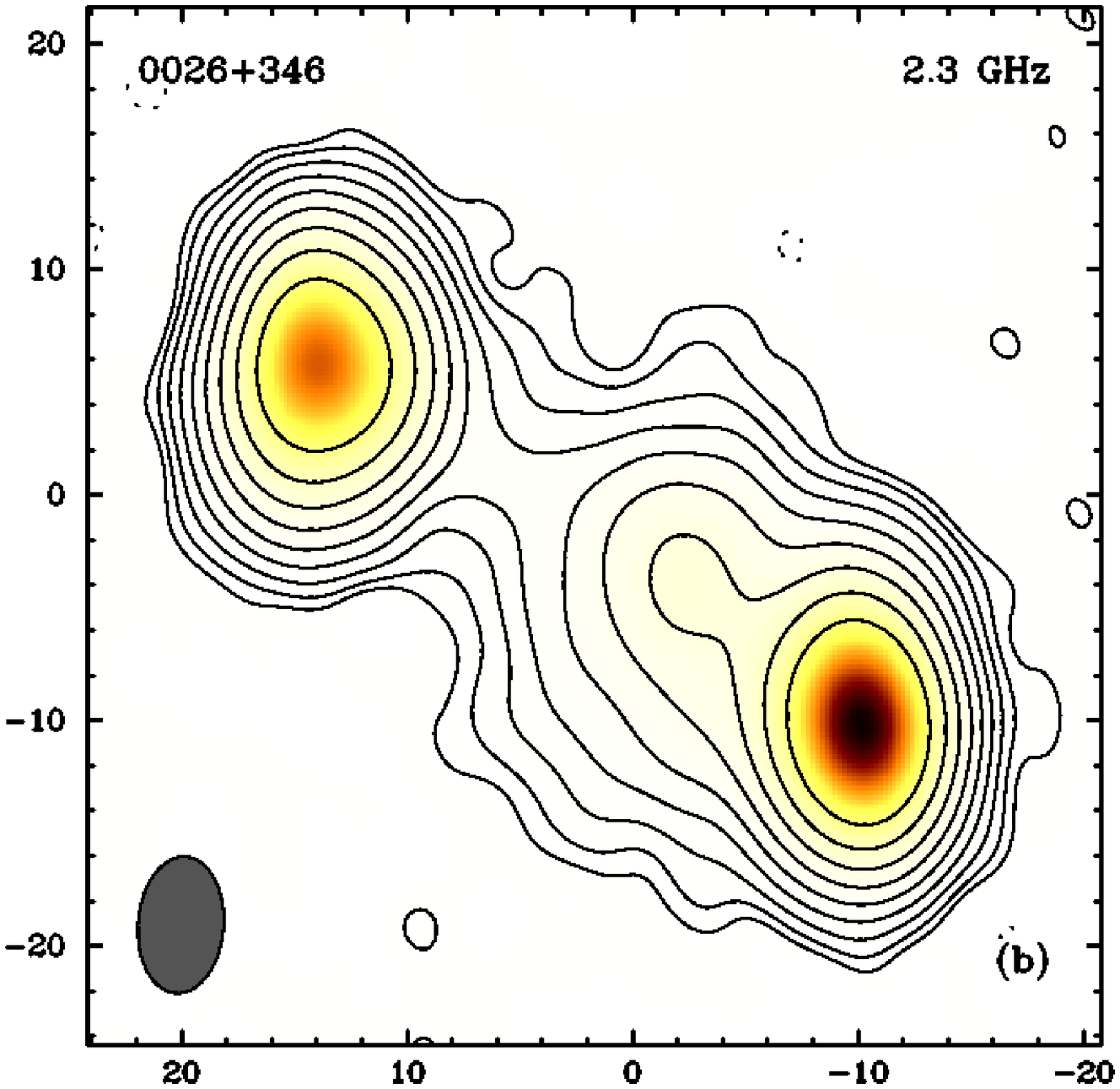}
\includegraphics{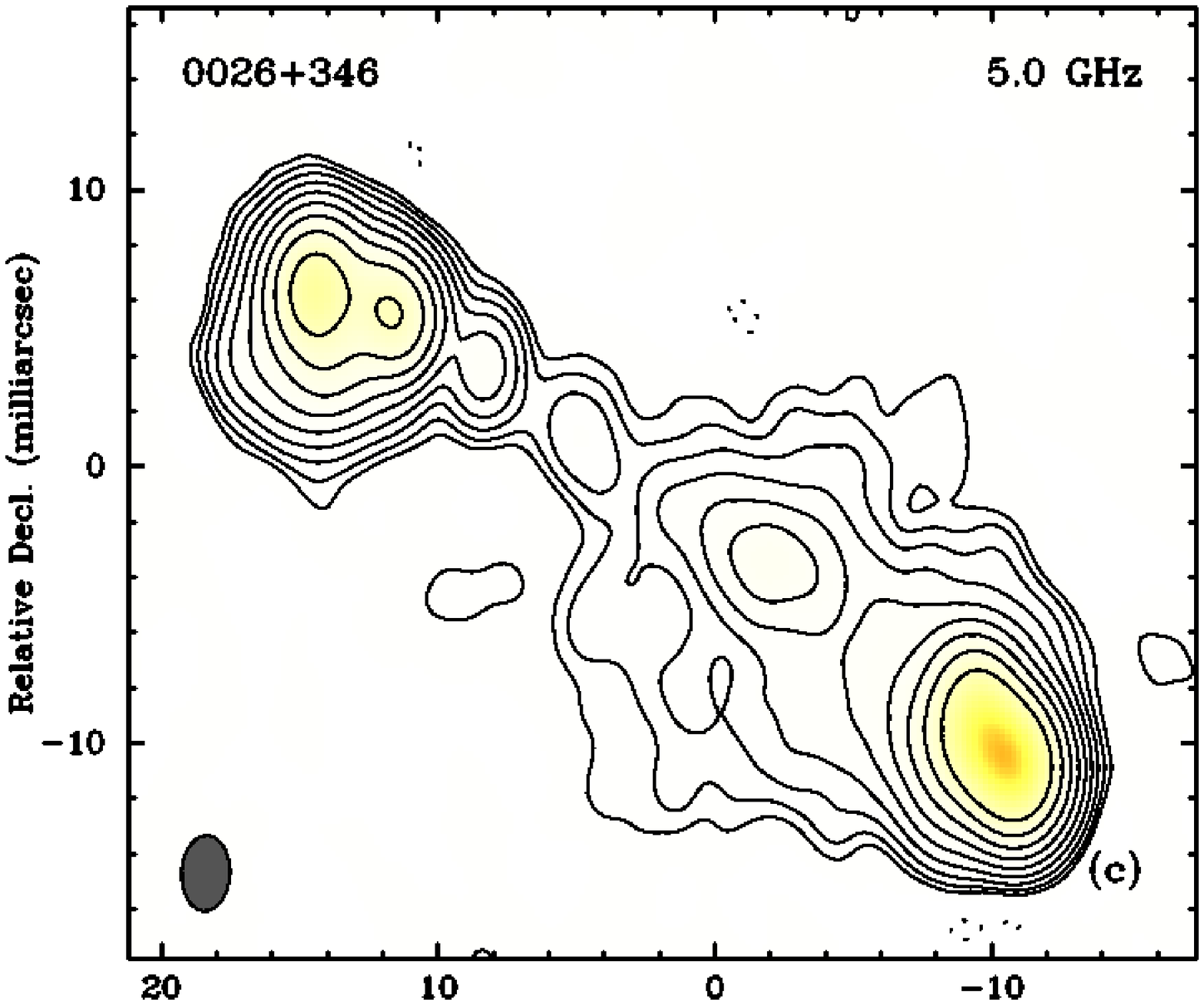}
\includegraphics{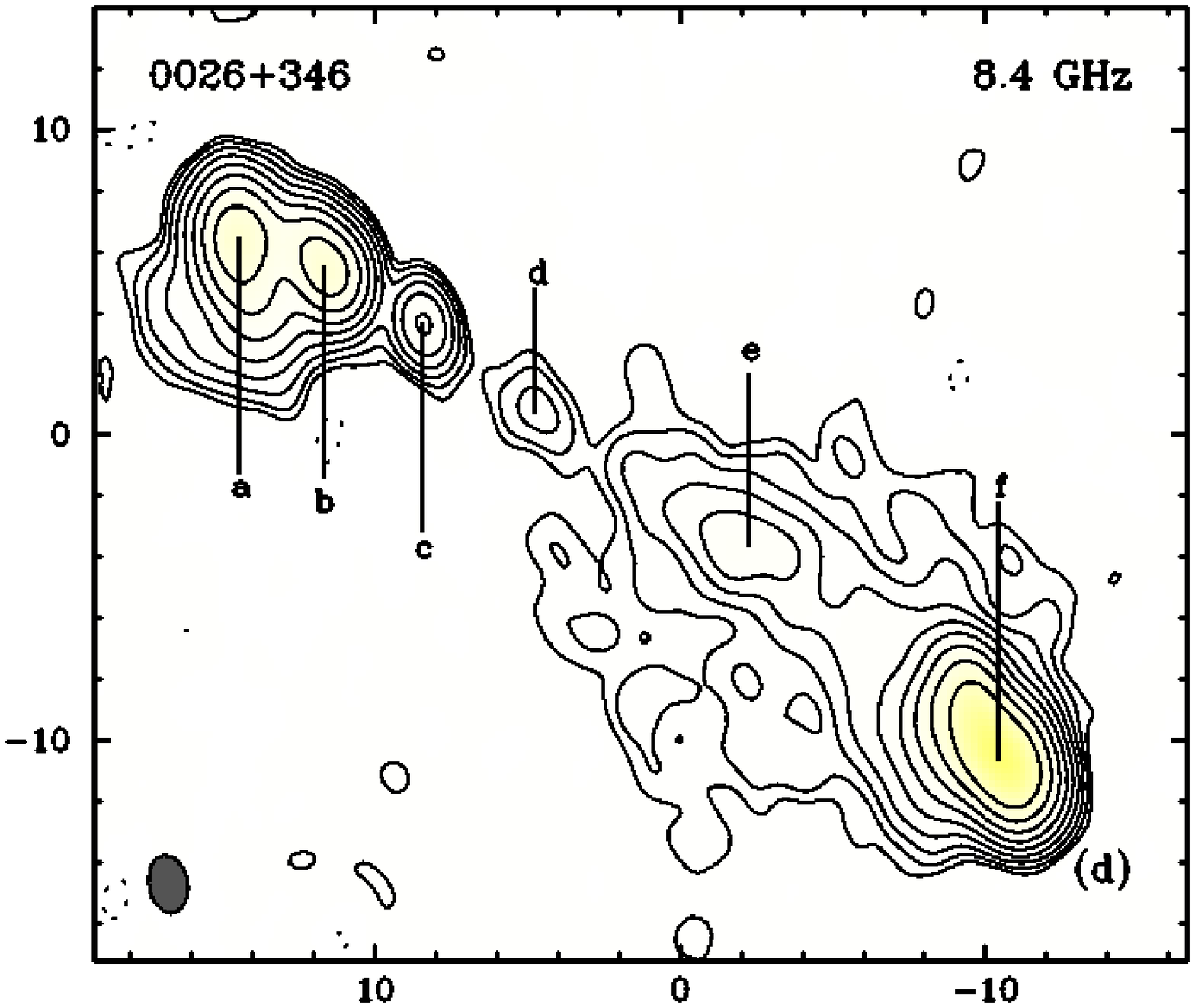}
\includegraphics{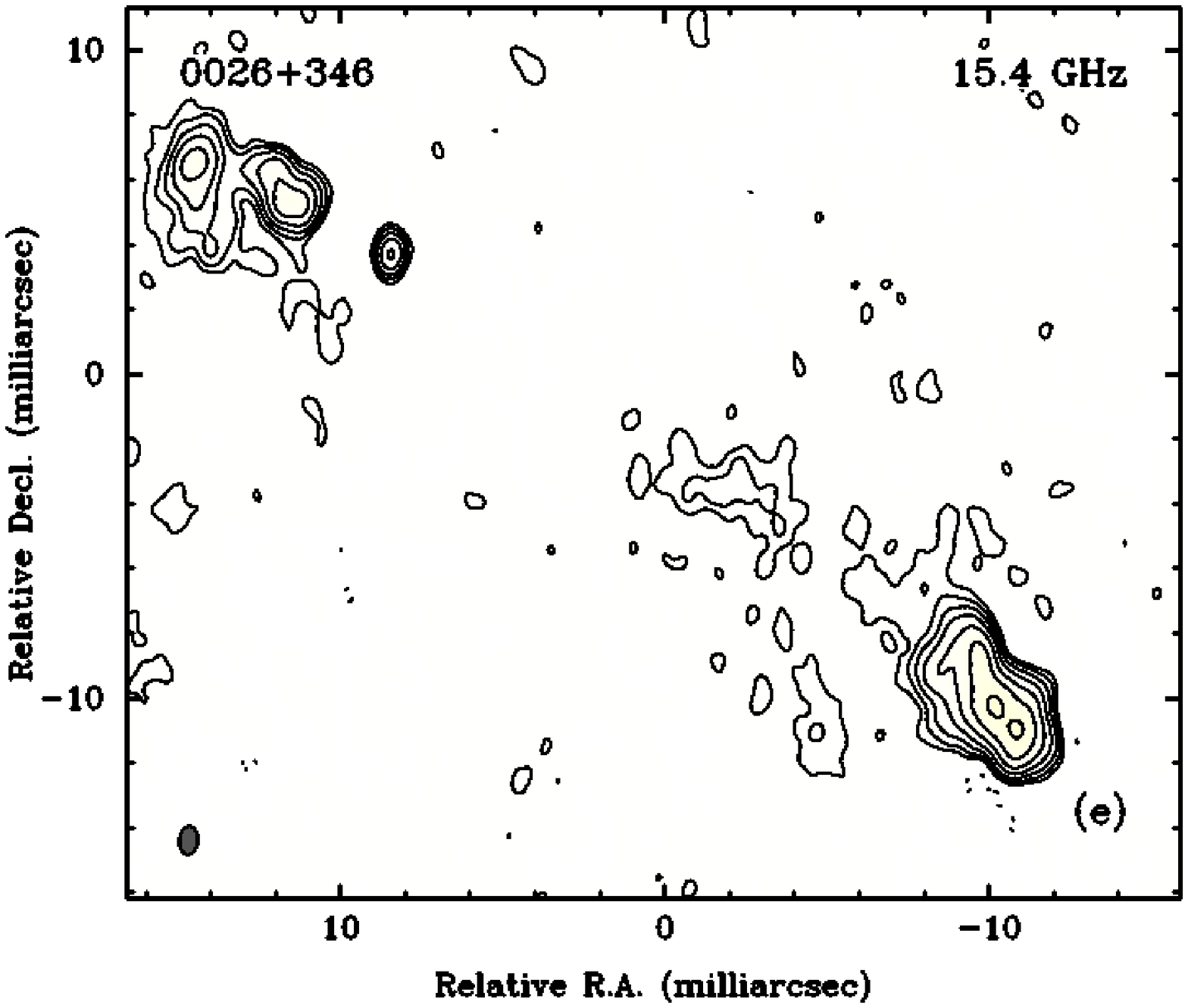}
\includegraphics{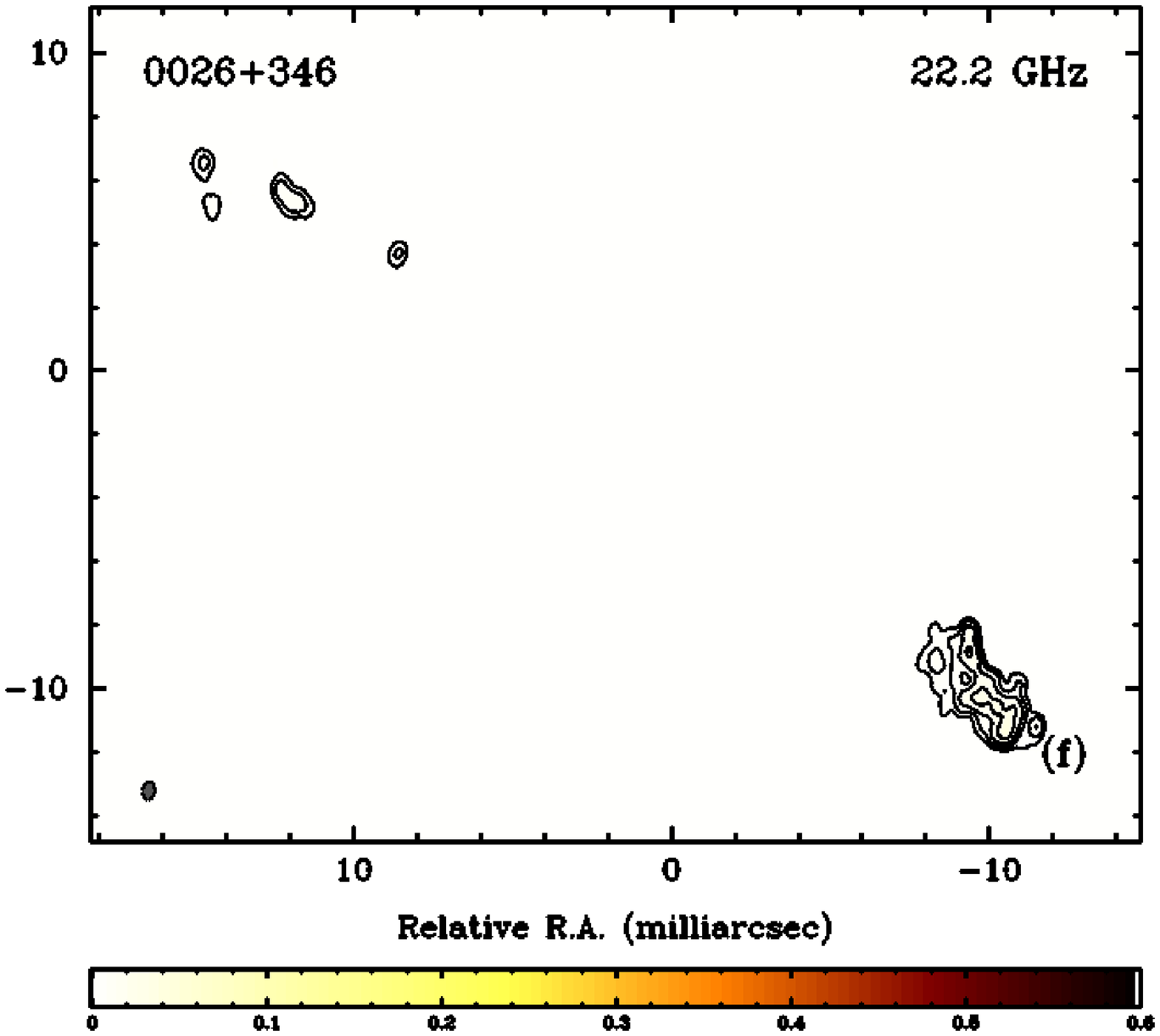}
\caption{CLEAN intensity maps of $J0029+3457$ at all frequencies, made with uniform weighting:
(a) 1.7 GHz; (b) 2.3 GHz; (c) 5.0 GHz; (d) 8.4 GHz; (e) 15.4 GHz; and (f) 22.2 GHz. 
The displayed contours represent $-$5, $-$2.5, 2.5, 5,
10, 20, 40,...  times the rms.  
The peak values and rms noise levels are:
(a) 0.534 \& 0.00017 Jy per beam; (b) 0.590 \& 0.00026 Jy per beam;
(c) 0.305 \& 0.00013 Jy per beam; (d) 0.187 \& 0.00011 Jy per beam;
(e) 0.054 \& 0.00030 Jy per beam; and (f) 0.0412 \& 0.00031 Jy per beam.  
The FWHM of the CLEAN beams (displayed in the lower left corners)
are:
(a) 6.81$\times$3.12 mas, @ $-$17.62$^\circ$; 
(b) 6.06$\times$3.80 mas, @ $-$4.37$^\circ$;
(c) 2.73$\times$1.74 mas, @ $-$2.61$^\circ$; 
(d) 1.89$\times$1.26 mas, @ 9.92$^\circ$;
(e) 0.87$\times$0.57 mas, @ $-$4.90$^\circ$; 
(f) 0.57$\times$0.37 mas, @ $-$6.32$^\circ$.
The maps are not all displayed to the same scale.}
\end{figure}

%Figure6
\begin{figure}
\phantom{mark}
\vspace{12.0cm}
\includegraphics{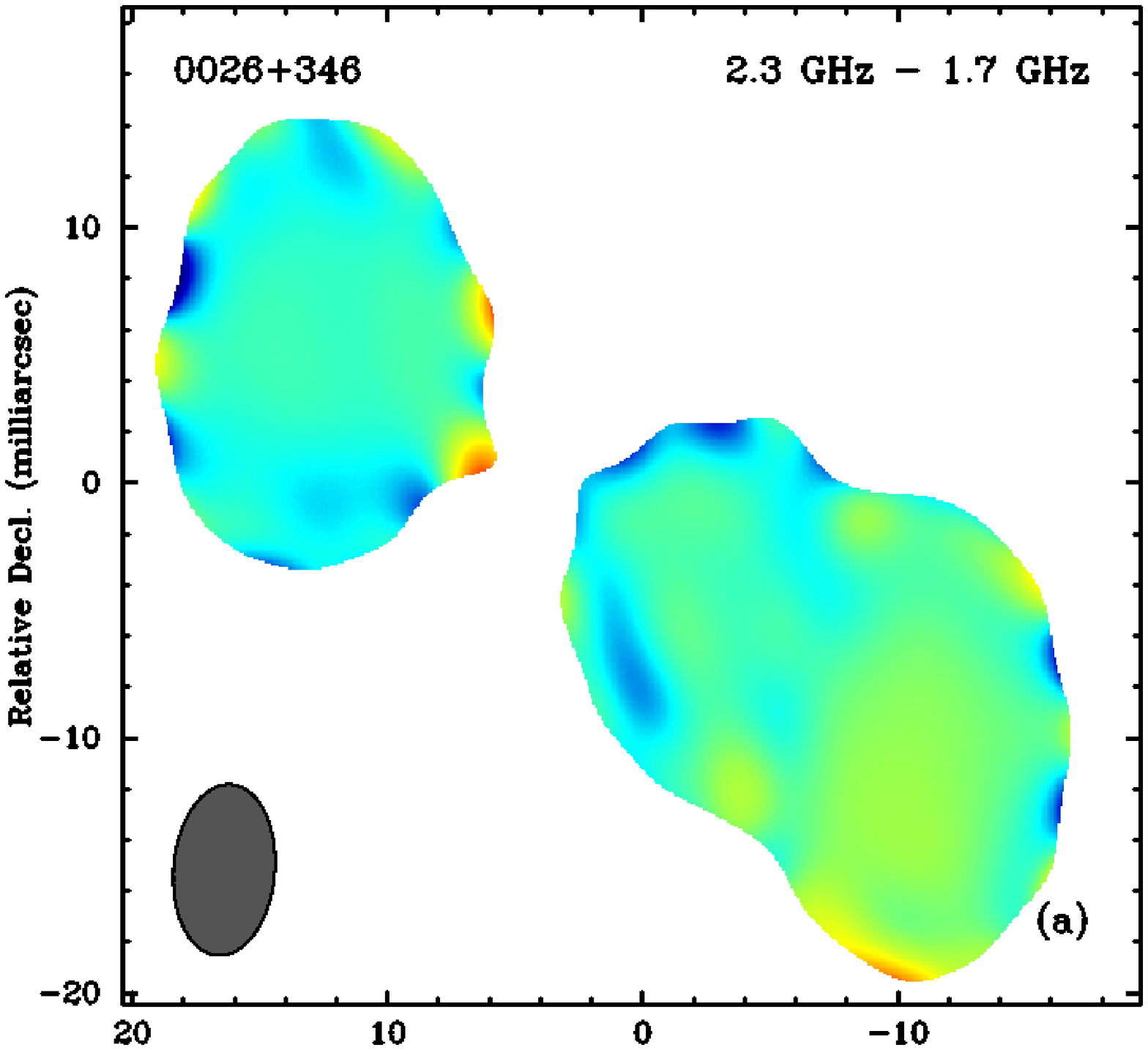}
\includegraphics{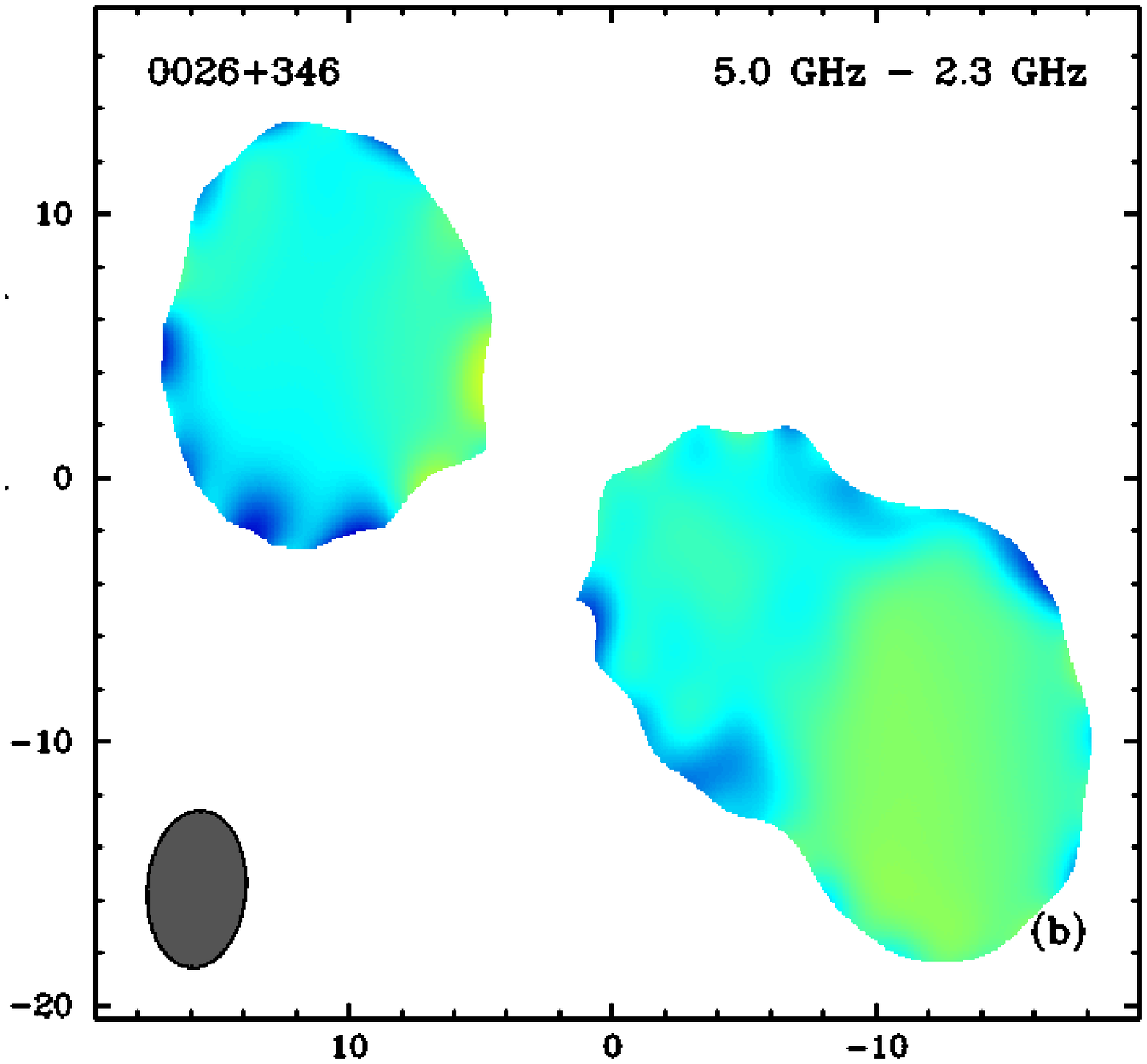}
\includegraphics{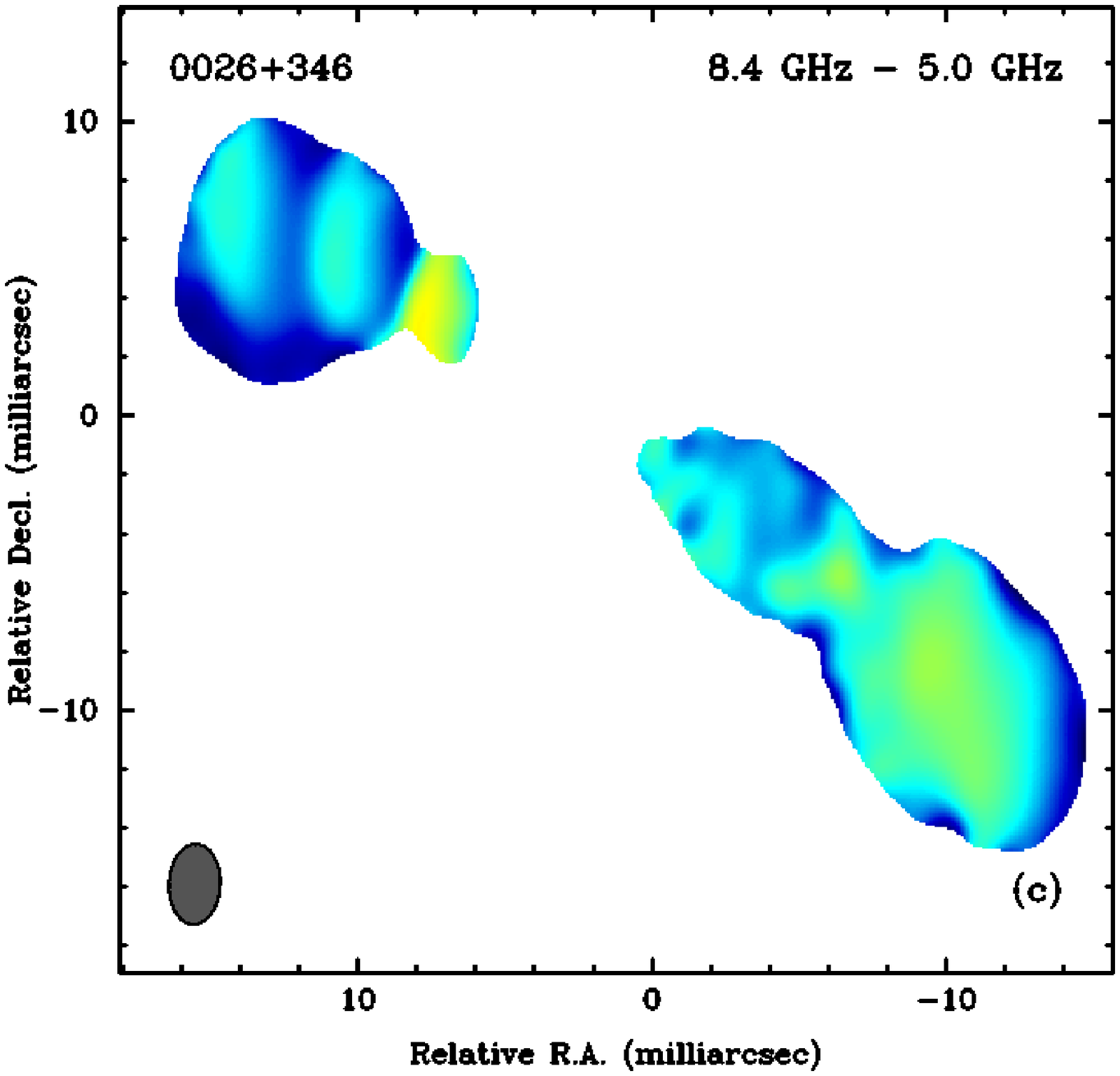}
\includegraphics{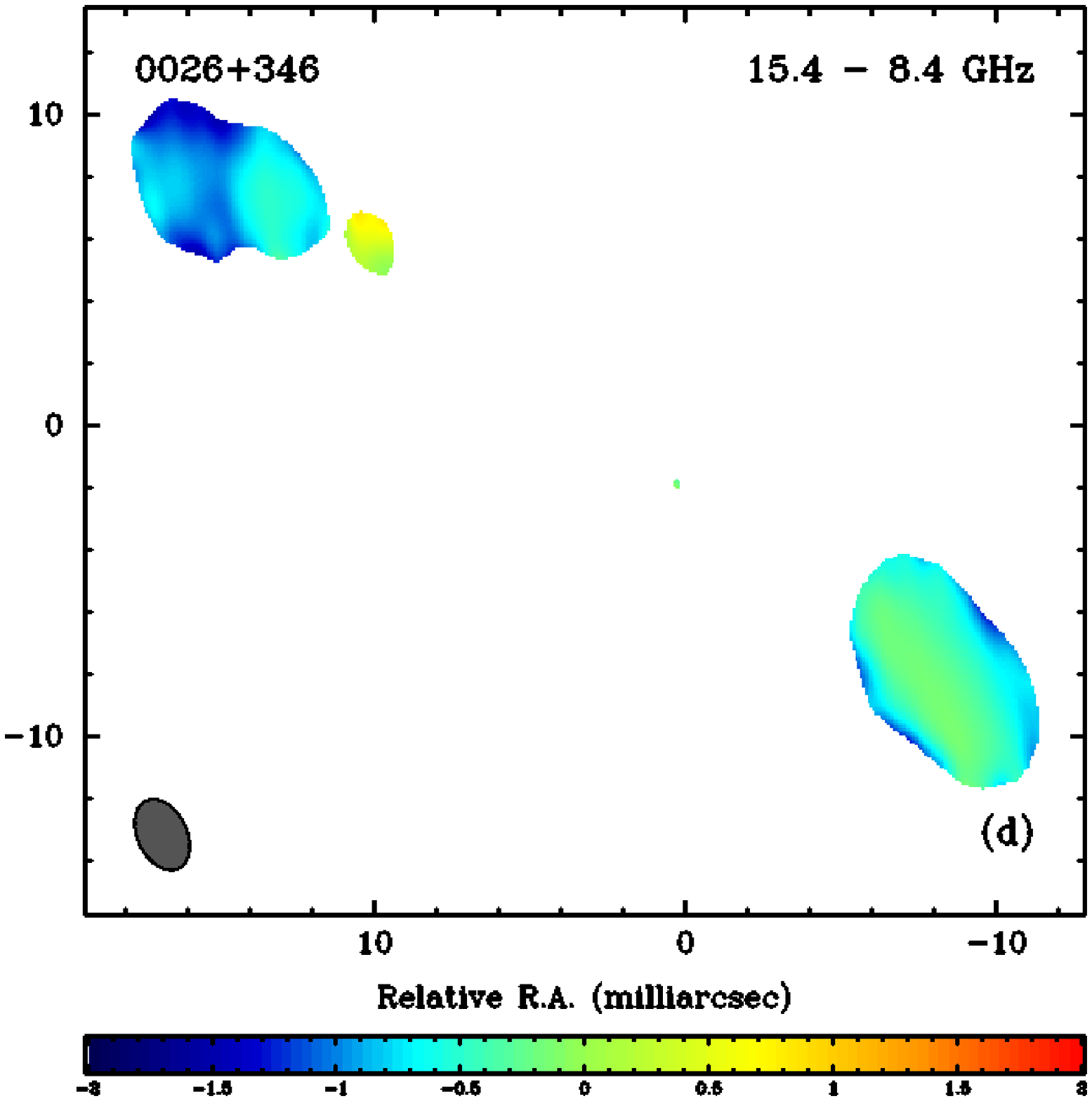}
\caption{Spectral-index maps of $J0029+3457$:
(a) between 2.3 and 1.7 GHz; (b) between 5.0 and 2.3 GHz;
(c) between 8.4 and 5.0 GHz; and 
(d) between 15.4 and 8.4 GHz.  
The colors indicate spectral index values, $\alpha$, as indicated on the wedge where
F$_\nu \propto \nu^\alpha$.  The bluer colors indicate more negative spectral indices (i.e. steeper, optically thin spectra). 
The uncertainties in the spectral indices are 
listed in Table 3.  The input CLEAN maps were both convolved with CLEAN beams displayed
in the lower left corners.  The maps are not all displayed to the same scale.}
\end{figure}

%Figure7
\begin{figure}
\vspace{6cm}
\includegraphics{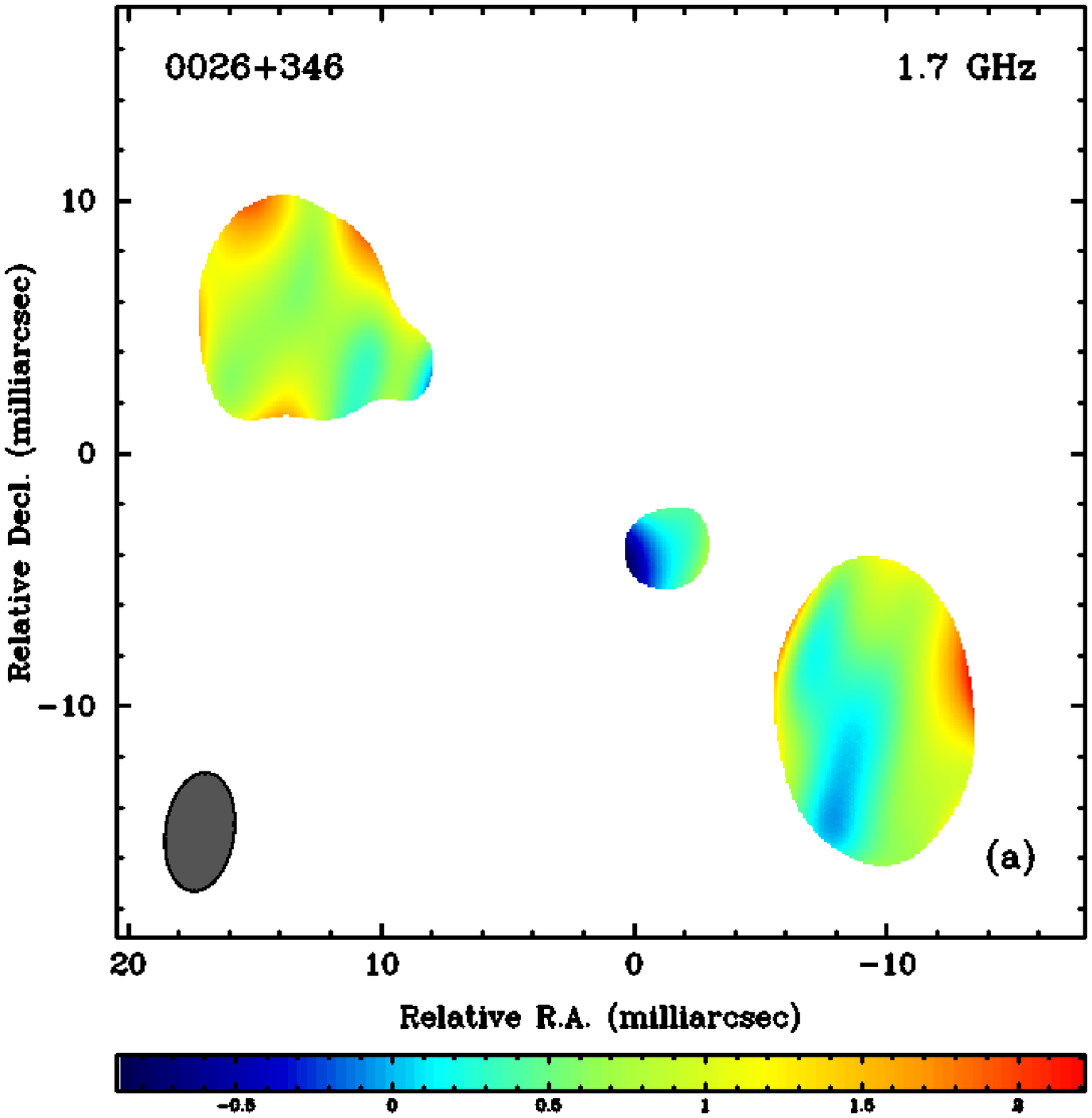}
\includegraphics{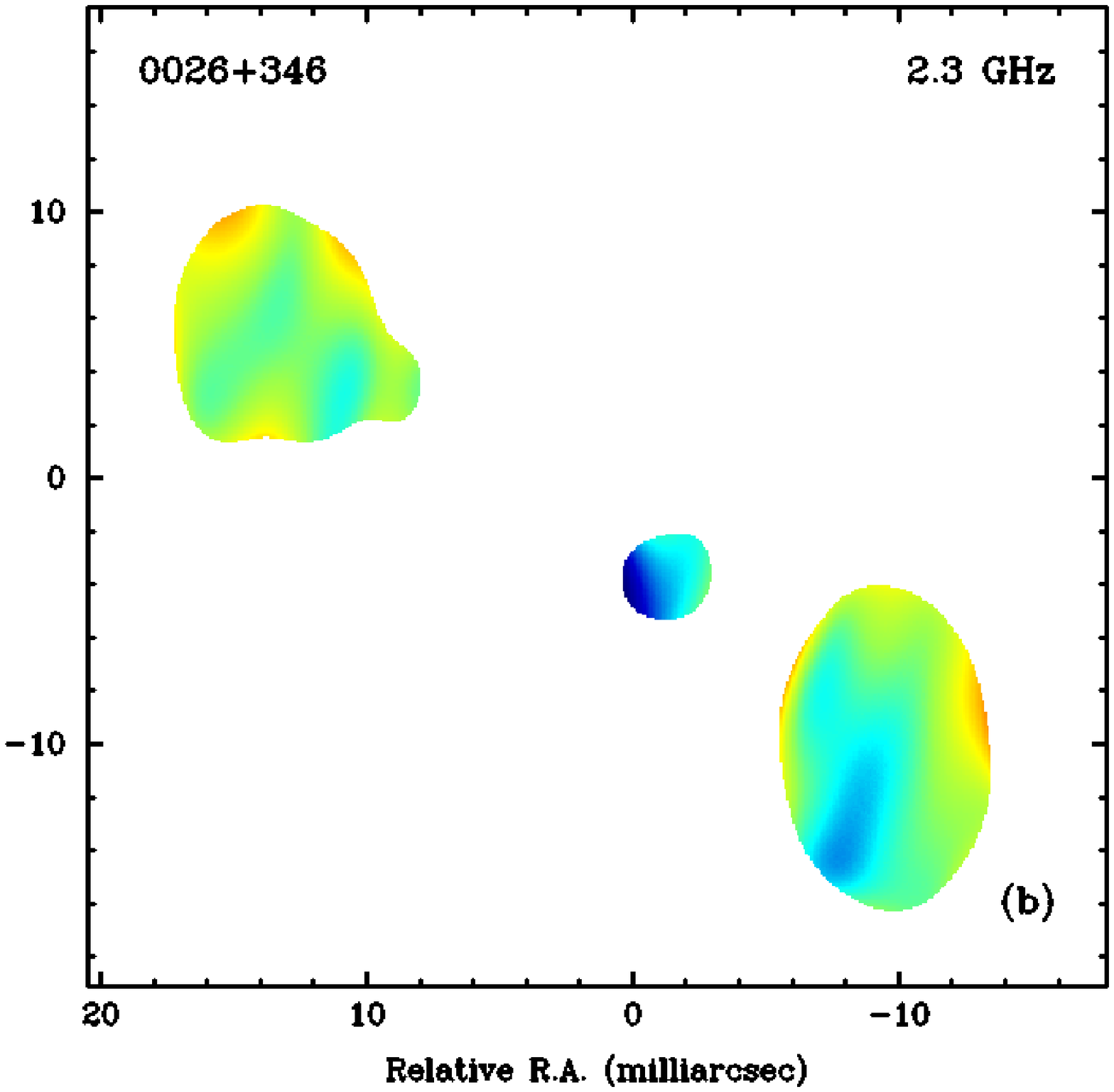}
\includegraphics{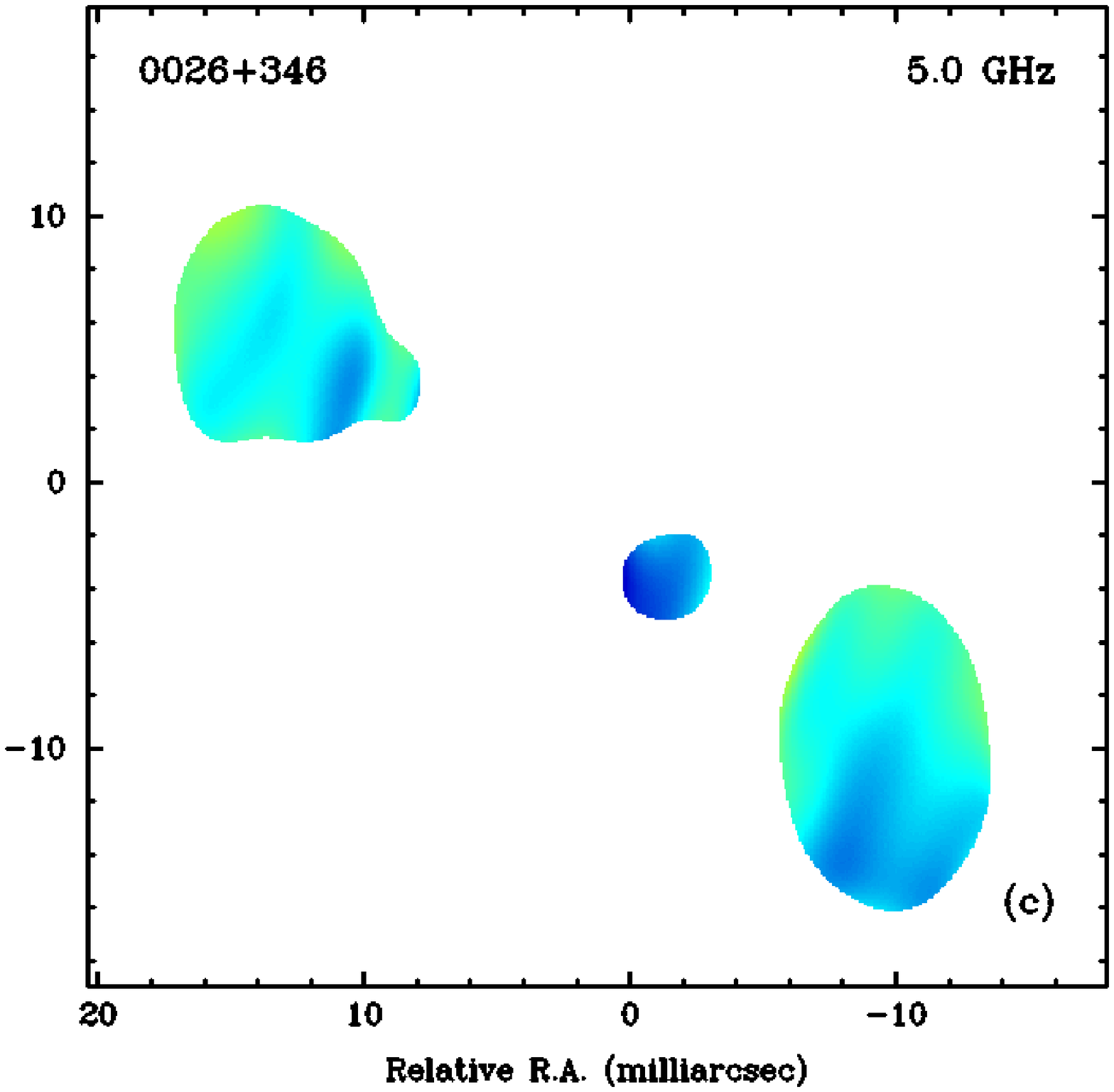}
\caption{Optical-depth maps of $J0029+3457$ at (a) 1.7 GHz; (b) 2.3 GHz; and (c) 5.0 GHz. The colors indicate optical depth values as indicated on the color wedge with redder colors indicating larger optical depths.  The uncertainties in the optical depths are listed in Table 3.}
\end{figure}

\bigskip
%Figure8
\begin{figure}
\vspace{5.5cm}
\includegraphics{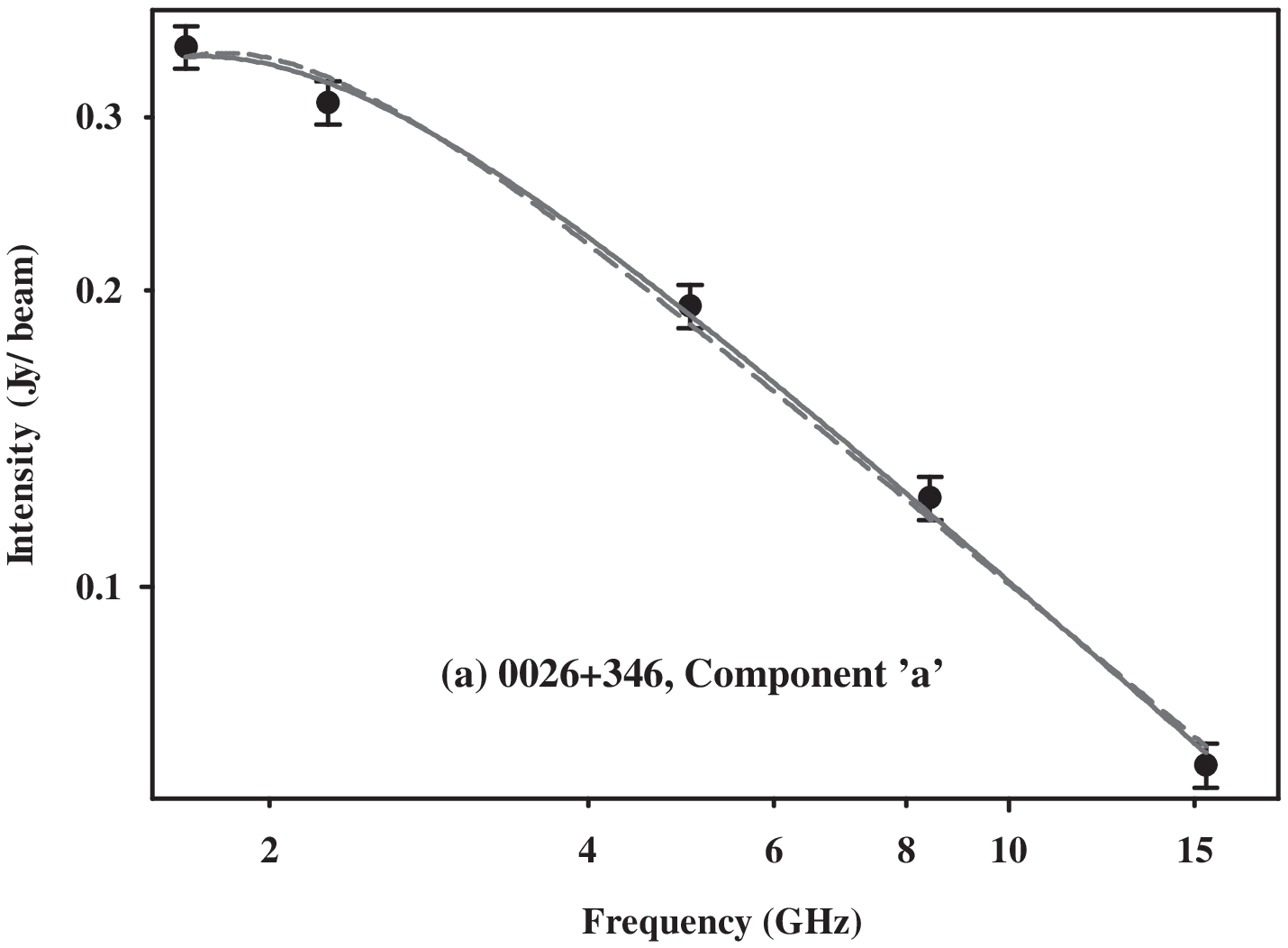}
\includegraphics{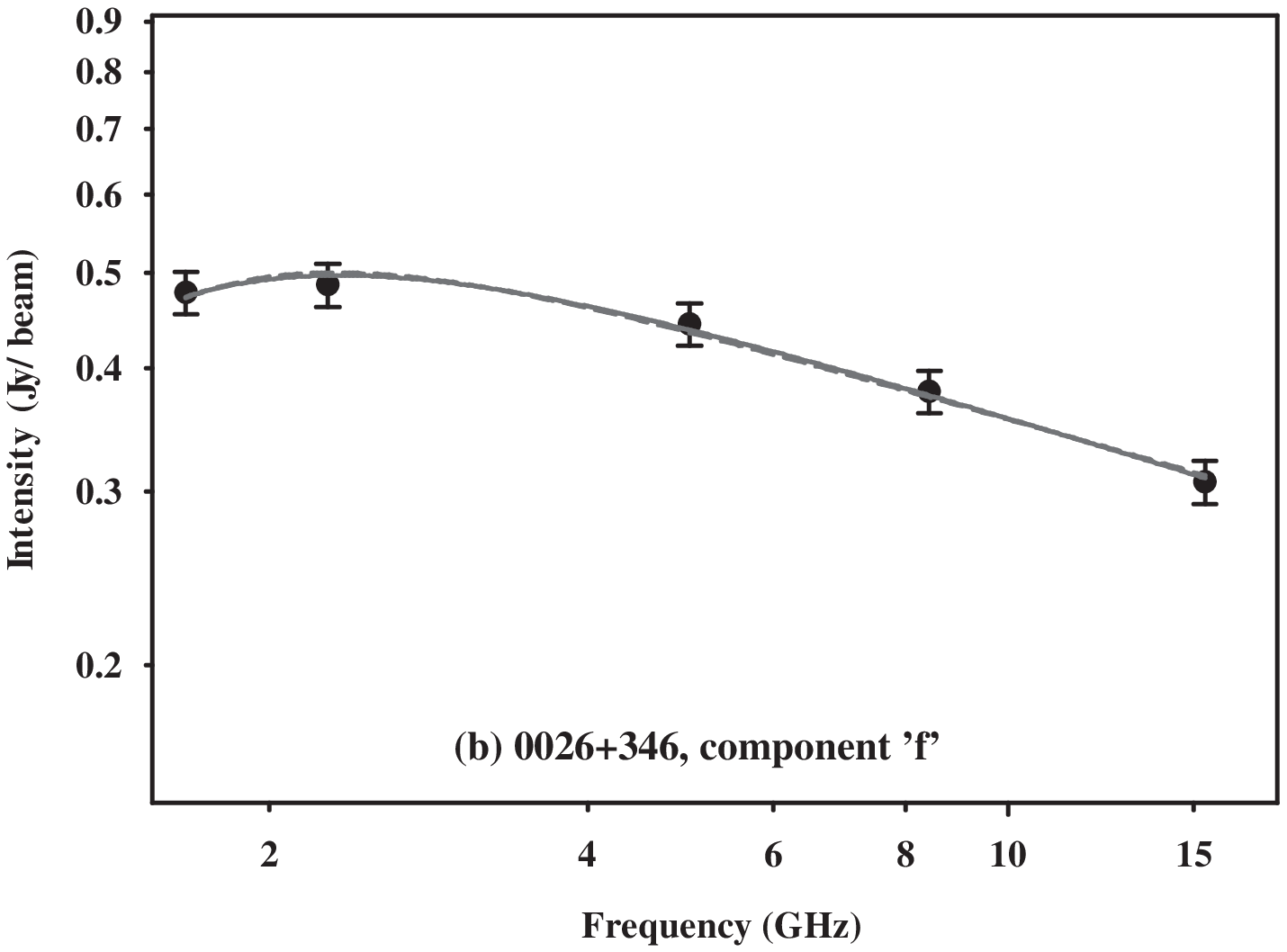}
\caption{Log-log plots of the spectra at the positions of the
(a) eastern peak and (b) western peak in emission in $J0029+3457$ at 5.0 GHz.  
The error bars represent the total intensity uncertainties, including
a 5\% uncertainty in the amplitude calibration.  The overlaid smooth lines are the curves of the best-fit models of a single FFA component (solid blue) and of a single SSA component (dashed red).}
\end{figure}

\clearpage
\section{Tables}

%Table 1
\begin{table}[htb]
\begin{center}
\caption{Specifics of observations.}
\medskip
\begin{tabular}{cccc}
\hline
Frequency & Antennas & Time On-Source & CLEAN map SNR \\
(GHz) & ~ & min &  \\
\hline
$J1324+4048$: \\
 1.663 & 123 & 120 & 1600 \\
 2.267 & 1 & 90 & 1800 \\
 4.983 & 12 & 90 & 970 \\
 8.417 & 12 & 90 & 690 \\
 15.359 & 12 & 108 & 260 \\
 22.233 & 12 & 158 & 250 \\
\hline
$J0029+3457$: \\
 1.663 & 123 & 106 & 3100 \\
 2.267 & 1 & 76 & 2300 \\
 4.983 & 12 & 76 & 2300 \\
 8.417 & 12 & 76 & 1700 \\
 15.359 & 12 & 106 & 180 \\
 22.233 & 12 & 136 & 130 \\
\hline
\tablenotetext{}{All observations were made with dual polarization
and a bandwidth of 512 MHz.}
\tablenotetext{}{Antennas:  1=all 10 VLBA antennas, 2=single VLA antenna,
3=Effelsberg.}
\end{tabular}
\end{center}
\end{table}

%\table 2
\begin{table}[htb]
\begin{center}
\caption{Polarization detection limits.}  
\medskip
\begin{tabular}{lcccc}
\hline
Frequency & Max.\ Linear Pol.\ & Max.\ Circular Pol.\ & Max.\ Linear Pol.\ & Max.\ Circular Pol.\ \\
(GHz) & (Jy per beam) & (Jy per beam) & (Jy per beam) & (Jy per beam) \\
\hline
~ & $J1324+4048$: & ~ & $J0029+3457$: & ~ \\
1.7 & 0.0076 & 0.00033 & 0.0044 & 0.00066 \\
2.3 & 0.0029 & 0.0011 & 0.0042 & 0.0013 \\
5.0 & 0.0011 & 0.00075 & 0.0013 & 0.00091 \\
8.4 & 0.00086 & 0.00060 & 0.0012 & 0.00090 \\
15.4 & 0.0023 & 0.0016 & 0.0019 & 0.0013 \\
\hline
\tablecomments{The values listed equal 3 $\times$ the rms
in each map.  No structure was seen in the raw polarization maps, and so any polarization
flux must be below these limits.}\end{tabular}
\end{center}
\end{table}

%table 3
\begin{table}[htb]
\begin{center}
\caption{Uncertainties in spectral-index maps and optical depth maps.}
\medskip 
\begin{tabular}{lcclc}
\hline
Freq. (GHz) & Min. $\Delta\alpha$ & Max. $\Delta\alpha$ & Freq (GHz) & $\Delta\tau$ \\
\hline
$J1324+4048$: & ~ & ~ & ~ & ~ \\ 
~~2.3 - 1.7  &  0.23 & 0.47 & ~~1.7  & 0.2 \\
~~5.0 - 2.3  &  0.090 & 0.18 & ~~2.3  & 0.16 \\
~~8.4 - 5.0  &  0.14 & 0.26 & ~~5.0  & 0.08 \\
~~15.4 - 8.4  & 0.12 & 0.26 & ~ & ~ \\
 $J0029+3457$: & ~ & ~ & ~ & ~ \\
~~2.3 - 1.7  &  0.23 & 0.49 & ~~1.7 & 0.2  \\
~~5.0 - 2.3  &  0.090 & 0.20 & ~~2.3 & 0.17 \\
~~8.4 - 5.0  &  0.14 & 0.29 & ~~5.0  & 0.09 \\
~~15.4 - 8.4  & 0.12 & 0.23 & ~ & ~ \\ 
\hline
\tablenotetext{}{Columns 2 lists the minimum uncertainties in $\alpha$, which correspond to locations 
across where the intensities are large; column 3 lists the maximum uncertainties, which occur at the edges of the source; column 5 lists the uncertainties in $\tau$, which are roughly constant across each source where the intensities are reasonably large.}
\end{tabular}
\end{center}
\end{table}

%table 4
\begin{table}[htb]
\begin{center}
\caption{Parameter values in the best $\chi_\nu^2$ fits to FFA and SSA models.}
\medskip
\begin{tabular}{crrrr}
\hline
Parameters & \multicolumn{2}{l}{\bf ~~$J1324+4048$} & \multicolumn{2}{l}{\bf ~~$J0029+3457$} \\
 ~ & East & West & 'a' & 'f' \\
\hline
FFA Model: \\
%\hline
I$_{2.3} {\rm GHz (Jy~bm^{-1})}$  & 0.505 & 0.396 & 0.420 & 0.588 \\
$\alpha$ & $-$1.19 & $-$0.927 & $-$0.950 & $-$0.333 \\
$\tau_{\rm FFA,0}$ & 4.00 & 3.87 & 1.43 & 0.939 \\
$\chi_{\nu}^2$ & 2.28 & 1.53 & 1.05  & 0.19 \\
\hline
SSA Model: \\
%\hline
I$_{2.3} {\rm (Jy~bm^{-1})}$ & 0.418 & 0.317 & 1.40 & 2.37 \\
$\alpha$ & $-$1.05 & $-$0.819 & $-$0.886 & $-$0.308 \\
$\tau_{\rm SSA,0}$ & 17.2 & 15.5 & 4.28 & 2.36 \\
$\chi_{\nu}^2$ & 5.74 & 3.47 & 2.03  & 0.34 \\
\hline
\end{tabular}
\tablenotetext{}{The models were fit to the observed intensities at the positions of the 5.0-GHz intensity peaks.}
\tablenotetext{}{The FFA model involves a homogeneous FFA component in front of a single homogeneous synchrotron radiation component and is given by 
$I_\nu = I_{2.3} ~ (\nu / 2.267) ^\alpha ~ {\rm exp}(-\tau_{\rm FFA,0} ~ \nu^{-2.1}).$
The SSA model involves a single homogeneous synchrotron source and is given by 
$I_\nu = I_{2.3}~(\nu / 2.267)^{2.5}~(1-{\rm exp}(-\tau_{\rm SSA,0}~\nu^{\alpha-2.5}))$.}
\end{center}
\end{table}

%table 5
\begin{table}[htb]
\begin{center}
\caption{Separation distances between features at different epochs and the inferred speeds.}
\medskip
\begin{tabular}{cccccccc}
\hline
Object & Freq.\ & Peaks & Epoch & Separation & 2004 Separation & Separation Rate \\
~ & (GHz) & ~&  yyyy-mm-dd & (mas) & (mas) & (mas yr$^{-1}$) \\
\hline
$J1324+4048$ & 4.98 & W-E & 1995-08-27 & 5.43 $\pm$ 0.16 & 5.44 $\pm$ 0.12 & 0.0004 $\pm$ 0.022 \\
~ & 15.4 & W-E & 2001-01-05 & 5.63 $\pm$ 0.11 & 5.59 $\pm$ 0.079 & $-$0.008 $\pm$ 0.038 \\
$J0029+3457$ & 4.98 & b-a & 1992-06-06 & 2.71 $\pm$ 0.36 & 2.803 $\pm$ 0.11 & 0.007 $\pm$ 0.031 \\
~ & ~ & ~ & 1994-09-15 & 2.73 $\pm$ 0.28 & ~ & 0.007 $\pm$ 0.031 \\
~ & ~ & f1-a & 1992-06-06 & 27.57 $\pm$ 0.29 & 28.404 $\pm$ 0.076 & 0.069 $\pm$ 0.025 \\
~ & ~ & ~ & 1994-09-15 & 27.759 $\pm$ 0.23 & ~ & 0.066 $\pm$ 0.020 \\
~ & ~ & f2-a & 1992-06-06 & 29.96 $\pm$ 0.23 & 30.250 $\pm$ 0.073 & 0.024 $\pm$ 0.020 \\
~ & ~ & ~ & 1994-09-15 & 30.049 $\pm$ 0.18 & ~ & 0.020 $\pm$ 0.016 \\
~ & 8.42 & b-a & 1995-03-22 & 2.80 $\pm$ 0.15 & 2.842 $\pm$ 0.083 & 0.004 $\pm$ 0.019 \\
~ & ~ & c-a & ~ & 6.49 $\pm$ 0.35 & 6.56 $\pm$ 0.19 & 0.008 $\pm$ 0.043 \\
~ & ~ & f1-a & ~ & 28.37 $\pm$ 0.11 & 28.522 $\pm$ 0.062 & 0.016 $\pm$ 0.013 \\
~ & ~ & f2-a & ~ & 30.355 $\pm$ 0.098 & 30.309 $\pm$ 0.059 & $-$0.005 $\pm$ 0.012 \\
\hline
\end{tabular}
\tablecomments{The `2004 Separation' correspond to the date of our observations on 2004-07-03.}
\end{center}
\end{table}

\end{document}